\documentclass[10pt]{revtex4}
\usepackage{amssymb,amsmath,amscd}
\usepackage{graphicx,calc,epsfig,pstricks,bbm}
\usepackage{tikz}
\newcommand {\be}{\begin{equation}}
\newcommand {\ee}{\end{equation}}
\newcommand{\ba}{\begin{array}{c}}
\newcommand{\ea}{\end{array}}
\newcommand{\group}[2]{\mathrm{#1}\left(#2\right)}
\renewcommand{\exp}[1]{\operatorname{exp}\left(#1\right)}

\newcommand{\tinysixj}[6]{\{\begin{smallmatrix}  #1&#2&#3\\#4&#5&#6\end{smallmatrix}\}}

\newcommand{\scr}{\scriptscriptstyle}
\newcommand{\vgraph}{\mathfrak{n}}
\newcommand{\cube}{\ba
 \begin{tikzpicture}
\pgfmathsetmacro{\cubex}{0.15}
\pgfmathsetmacro{\cubey}{0.15}
\pgfmathsetmacro{\cubez}{0.15}
\draw (0,0,0) -- ++(-\cubex,0,0) -- ++(0,-\cubey,0) -- ++(\cubex,0,0) -- cycle;
\draw (0,0,0) -- ++(0,0,-\cubez) -- ++(0,-\cubey,0) -- ++(0,0,\cubez) -- cycle;
\draw (0,0,0) -- ++(-\cubex,0,0) -- ++(0,0,-\cubez) -- ++(\cubex,0,0) -- cycle;
\end{tikzpicture}
\ea}
\newcommand{\parallelsum}{\mathbin{\!/\mkern-5mu/\!}}

\begin{document}

\title{Quantum-Reduced Loop Gravity: Cosmology}
\author{Emanuele Alesci}
\email{emanuele.alesci@fuw.edu.pl}
\affiliation{Institute for Quantum Gravity, FAU Erlangen-N\"urnberg, Staudtstr. 7, D-91058 Erlangen, Germany, EU}
\affiliation{Instytut Fizyki Teoretycznej, Uniwersytet Warszawski, ul. Ho{\.z}a 69, 00-681 Warszawa, Poland, EU}
\author{Francesco Cianfrani}
\email{francesco.cianfrani@ift.uni.wroc.pl}
\affiliation{Instytut Fizyki Teoretycznej, Uniwersytet Wroc\l awski, pl. M. Borna 9, 50-204 Wroc\l aw, Poland, EU.}

\begin{abstract}
We introduce a new framework for loop quantum gravity: mimicking the spinfoam quantization procedure we propose to study the symmetric sectors of the theory imposing the reduction weakly on the full kinematical Hilbert space of the canonical theory.  As a first application of Quantum-Reduced Loop Gravity we study the inhomogeneous Bianchi I model. The emerging quantum cosmological model represents a simplified arena on which the complete canonical quantization program can be tested. The achievements of this analysis could elucidate the relationship between Loop Quantum Cosmology and the full theory.
\end{abstract}

\maketitle

\section{Introduction}
The realization of a quantum theory for the gravitational field must provide an explanation to the current puzzles of General Relativity (GR), {\it i.e.} the presence of mathematical singularities. These singularities have been shown to be unavoidable in some symmetry reduced models describing relevant physical situations, such as the collapse of standard matter and the beginning (eventually also the end) of the Universe evolution \cite{sing}. Hence, it is demanded to a quantum formulation of gravity to answer to the questions posed by the unpredictability of GR in these cases. 

Loop Quantum Gravity (LQG) \cite{revloop,thiemann book} constitutes the most advanced model which pursues the quantization of geometric degrees of freedom. It is based on a canonical quantization {\it a la Dirac} of the holonomy-flux algebra associated with Ashtekar-Barbero variables \cite{ABvar} in the Hilbert space of distributional connections. One first defines a kinematical Hilbert space in which the Gauss constraint is then solved. The resulting basis elements are the so-called spinnetworks: these are labeled by graphs $\Gamma$ and belong to $\mathcal{L}^2(SU(2)^E / SU(2)^V)$, $E$ and $V$ being the total number of edges and vertexes of $\Gamma$, respectively. The invariance under diffeomorphisms is then implemented by summing over the orbit of the associated operator, which gives the so-called s-knots \cite{sknots}: these are distributional states representing the equivalence class of spinnetworks under diffeomorphisms. In the space of s-knots the superHamiltonian operator can be regularized \cite{qsd,ar} and thanks to diff-invariance the regulator can be safely removed leading to an anomaly free quantization of the Dirac algebra. However, particularly in view of the presence of the volume operator \cite{vol1,vol2}, the explicit analytical expression for the matrix elements of the superHamiltonian and the properties of the physical Hilbert space are still elusive. For these reasons other approaches such as the master constraint program \cite{mc} or the more recent deparametrized system in terms of matter fields \cite{dep} have been introduced in the canonical framework.

Cosmology is a natural arena to test the theory and its dynamics due to the high degree of symmetry of the configuration space. The cosmological implementation of LQG has been realized in the framework of Loop Quantum Cosmology (LQC) \cite{revlqc1,revlqc2} (see \cite{rv,brv,liv} for alternative proposals). This is based on the implementation of a minisuperspace quantization scheme, in which the phase space is reduced on a classical level according with the symmetries of the model. Because the Universe is described by a homogeneous (and eventually isotropic) space-time manifold, the resulting configuration space is parametrized by three spatial-independent variables. These variables describe the connections and the momenta of the reduced model after a gauge-fixing of both the $SU(2)$ gauge symmetry and diffeomorphisms invariance has been performed. As a consequence, the regularization of the superHamiltonian operator can be accomplished by fixing an external parameter $\bar\mu$ related with the existence of an underlying quantum geometry \cite{qnb} (see \cite{HE} for a critical discussion on the regularization in LQC). The resulting theory is a well established research field with several remarkable features  and physical consequences, the main ones being a bounce replacing the initial singularity \cite{qnb,bohe,ashedI,mm,ashedII}, the generation of initial conditions for inflation to start \cite{AS,LB} and the prediction of peculiar effects on the cosmic microwave background radiation spectrum \cite{M08,M09,MCGB,GBCM,BCT,AAN} (see also \cite{CBGV,MCBG,CMBG}). 

However LQC has not yet been shown to be the cosmological sector of LQG and in order to solve the tension between the regularization procedures of the two theories, new approaches have been recently envisaged in order to provide an alternative definition of the superHamiltonian operator in the full theory (see \cite{bola} that bring it closer to the $\bar{\mu}$ scheme of LQC).
In this paper, we give a detailed presentation of the procedure introduced in \cite{noi}, in which we adopt the opposite view-point assuming LQG as the correct theory obtained by quantizing GR and then we look for its cosmological sector imposing a symmetry-reduction at the quantum level. This way we construct a theory in which we first quantize and then reduce instead of first classically reducing and then quantizing as it's usually done in LQC.  This approach is not expected to work only in cosmology, but it can be extended also to other symmetric sectors of the theory. This way, we define a new framework for the analysis of the implications of LQG in relevant (symmetry-reduced) physical cases (\emph{Quantum-reduced Loop Gravity}).
Our cosmological quantum model will then be a proper truncation of the full kinematical Hilbert space of LQG. The virtue of our approach mainly consists in the possibility to realize a fundamental description of a cosmological space-time, which fills the gap with the full theory and on which Thiemann's regularization procedure for the superHamiltonian \cite{qsd} can be applied.

The paper is organized as follows:

In sections \ref{lqgintroduction} we quickly review the main tools of the LQG quantization of GR, while in section \ref{Bianchiom} the homogeneous Bianchi models are presented and the LQC framework is shortly discussed. 
Then in section \ref{BianchiI}  we perform a classical analysis and we outline how, by considering a proper inhomogeneous extension, it is possible to retain a certain dependence from spatial coordinates into the reduced variables describing a Bianchi I model. Within this scheme, we get the following set of additional symmetries: i) three independent $U(1)$ gauge transformation, denoted by $U(1)_i$ ($i=1,2,3$),defined in the 1-dimensional space generated by fiducial vectors $\omega_i=\partial_i$, and ii) reduced diffeomorphisms, which act as 1-dimensional diffeomorphisms along a given fiducial direct $i$ and rigid translations along the other directions $j\neq i$. We also outline how a similar formulation will be relevant within the BKL conjecture \cite{BKL} scheme.

In section \ref{reduced} we discuss the implications of this formulation in a reduced quantization scheme. The elements of the associated Hilbert space are defined over {\emph reduced graphs}, whose edges are parallel to fiducial vectors and to each edge $e_i\parallelsum\partial_i$ is associated a $U(1)_i$ group element. Within this scheme, a proper quantum implementation can be given to the algebra of reduced holonomy-flux variables. The additional symmetries can then be  implemented as in full LQG and they imply the conservation of $U(1)_i$ quantum numbers along the integral curves of fiducial vectors $\partial_i$ and that states have to be defined over reduced s-knots. However, we will note that no meaningful expression for the superHamiltonian operator can be given. 

The failure of reduced quantization to account for the proper dynamics is the motivation for considering a different approach, in which a truncation of full LQG is performed.
This is done in section \ref{CLQG} where the truncation is realized such that 
\begin{enumerate} \item the elements of the full Hilbert space are defined over reduced graph:
 this is implemented via a projection and this implies the restriction of arbitrary diffeomorphisms to reduced ones.
\item The SU(2) gauge group is broken to the $U(1)_i$ subgroups along each edge $e_i$:
 this is realized by imposing weakly a gauge-fixing condition on each group element over an edge $e_i$.
\end{enumerate}
 A proper quantum-reduced kinematical Hilbert space is found by mimicking the analogous procedure adopted in Spin-Foam models to solve the simplicity constraints \cite{sfsimpl}. In particular, we develop projected $U(1)_i$-networks \cite{dl} by which we can embed functionals over the $U(1)_i$ group into functionals over the $SU(2)$ group. Hence, we impose strongly a Master constraint condition obtained by squaring and summing all the gauge-fixing conditions. This requirement fixes the relation between $SU(2)$ and $U(1)_i$ quantum numbers and the resulting projected $U(1)_i$ networks solve the gauge-fixing conditions weakly. At the end, the reduced $U(1)_i$ elements are obtained from full $SU(2)$ ones by projecting over the states with maximum magnetic number along the internal direction $i$. 
The projection to $U(1)_i$ elements can then be applied directly to $SU(2)$-invariant states. As a result some non-trivial intertwiners are induced between $U(1)_i$ group elements for different values of the index $i$. These intertwiners coincide with the projection of the coherent Livine-Speziale intertwiners \cite{ls} on the usual intertwiners base. Hence, the $U(1)_i$ states are not kinematically independent, but they realize a true three-dimensional vertex structure. 
This result allows us to implement the superHamiltonian operator according with Thiemann regularization scheme \cite{qsd}. In fact, by defining states over reduced s-knots it is possible to remove the regulator and get a well-defined expression. Moreover, thanks to the simplifications due to the reduced Hilbert space structure (the volume operator is diagonal!), we evaluate in section \ref{Hmatrix} the explicit expression of the superHamiltonian matrix elements in the case of a 3-valence vertex. Concluding remarks follow in section \ref{concl}.

\section{Loop Quantum Gravity}\label{lqgintroduction} 
The kinematical Hilbert space of LQG $\mathcal{H}^{kin}$ is developed by quantizing the holonomy-flux algebra of the corresponding classical model, whose phase space is parametrized by Ashtekar-Barbero connections $A^i_a$ and densitized triads $E^a_i$. In particular, the space of all holonomies is embedded into the space of generic homomorphisms from the set of all piecewise analytical paths of the spatial manifold into the topological SU(2) group $\bar{X}$ \cite{ALMMT95}. On such a space a regular Borel probability measure is induced from the SU(2) Haar one and the kinematical Hilbert space for a graph $\Gamma$ is the tensor product of $\mathcal{L}^2(\bar{X},d\mu)$ for each edge $e$. A basis in this kinematical Hilbert space can be obtained using the Peter-Weyl theorem. Introducing an $ SU(2)$ matrix element in representation $j$, $\langle g | j,{\alpha,\beta}\rangle=D^{j}_{\alpha\beta}(g)$, the generic basis element of $H^{kin}_{\Gamma}$ for a given graph $\Gamma$ with edges $e$ will be of the form:
\be
\langle h_e|\Gamma, j_e, {\alpha_e,\beta_e}\rangle=  \bigotimes_{e\in\Gamma} D^{j_e}_{\alpha_e\beta_e}(h_e),
\label{spinnetwork}
\ee
from which we can reconstruct the whole kinematical Hilbert space as $\mathcal{H}^{kin}=\bigoplus_{\Gamma} \mathcal{H}^{kin}_{\Gamma}$.


 Fluxes $E_i(S)$ across a surface $S$ are quantized such that a faithful representation of the holonomy-flux algebra is realized and they turn out to act as left (right)-invariant vector fields of the SU(2) group. In particular, given a surface $S$ which intersects $\Gamma$ in a single point $P$ belonging to an edge $e$ such that $e=e_1\bigcup e_2$ and $e_1\cap e_2=P$, 
the action of $\hat{E}_i(S)$ reads
\be
\hat{E}_i(S)D^{(j_e)}(h_e)
=8\pi\gamma l_P^2 \; o(e,S) \; D^{j_e}(h_{e_1})\,^{j_e}\tau^{i}\,D^{j_e}(h_{e_2}).\label{Eop}
\ee  
$\gamma$ and $l_P$ being the Immirzi parameter and the Planck length, respectively, and the factor $o(e,S)$  is equal to $0,1,-1$ according to the relative sign of $e$ and the normal to $S$, while $^{j_e}\tau^i$ denotes the SU(2) generator in $j_e$-dimensional representation.  
  
The set of GR constraints in Ashtekar variables, {\it i.e.} the Gauss constraint  $\mathcal{G}$,  generating SU(2) gauge symmetry, the vector constraint $V_a$, generating 3-diffeomorphisms, and the Hamiltonian constraint $H$, generating time-reparametrizations, are implemented in $\mathcal{H}^{kin}$ according with the Dirac prescription for the quantization of constrained systems \cite{ht}, namely promoting the constraints to operators acting on $\mathcal{H}^{kin}$ and looking for the Physical Hilbert space $\mathcal{H}^{Phys}$ where the operator equations  $\hat{\mathcal{G}}=0, \hat{V_a}=0, \hat{H}=0$ hold.
We quickly review how these constraints are implemented in LQG:

\begin{itemize}

\item
$\mathcal{G}$ maps $h_e$ in $h_e'=\lambda_{s(e)}h_e\lambda^{-1}_{t(e)}$, $s(e)$ and $t(e)$ being the initial and final points of $e$, respectively, while $\lambda$ denotes $SU(2)$ group elements and  the condition $\mathcal{G}=0$ is solved implementing a group averaging procedure.  To this aim one introduces a projector $P_{\mathcal{ G}}$ to the $SU(2)$-invariant Hilbert space $^{\mathcal{G}}\mathcal{H}^{kin}$, by integrating over the $SU(2)$ group elements $\lambda_{s(e)}$ and $\lambda_{t(e)}$ for each edge.
Basis elements of $^{\mathcal{G}}\mathcal{H}^{kin}$ are then the so called  {\it spinnetworks}:
 \be
< h |\Gamma,\{j_e\},\{x_v\}>=\prod_{v\in\Gamma} \prod_{e\in\Gamma} {x_{v}}\cdot D^{j_{e}}(h_{e}),  
\label{spinnet}
\ee

$x_{v}$ being the $SU(2)$ invariant intertwiners at the nodes $v$ and they can be seen as maps between the representations associated with the edges emanating from $v$ and $\cdot$ means index contraction.

\item
The action of finite diffeomorphisms $\varphi$ maps the original holonomy into the one evaluated on the transformed path, $h_e\rightarrow h_{\varphi(e)}$: states invariant under this action can be found in the dual of $\mathcal{H}^{kin}$ and they are the so-called {\it s-knots} \cite{sknots}, namely equivalence class of spinnetworks under diffeomorphisms.

\item 
The hamiltonian constraint $\hat{H}$  in the gauge and diffeomorphisms invariant Hilbert space can be regularized by adopting the standard prescription given by Thiemann \cite{qsd} or an alternative recent proposal \cite{ar}, but at present only the first one has been shown to reproduce the Dirac algebra without anomalies.
We resume Thiemann construction because it will be adapted to the cosmological model of interest in this article.

We restrict our attention to the so-called Euclidean part of the Hamiltonian constraint, which can be written as
 \begin{align}
\begin{split}
  H[N]=  
       \int_\Sigma d^3x \, N(x)\,H(x)
     =  - 2 \int_\Sigma  \, N  \ {\rm Tr}(F \wedge   \{ A, V \}),             
  \label{H_E_N0}  
 \end{split}
 \end{align}
$V$ being the volume operator of the full space, while $A$ and $F$ denote the connection 1-form and the curvature 2-form, respectively.
The regularization is based on defining a triangulation $T$ adapted to the graph $\Gamma$ on which the operator acts. In particular, for each pair of links $e_i$ and $e_j$ incident at a node $v$ of  $\Gamma$ we choose a semi-analytic arcs $a_{ij}$  whose end points $s_{e_i},s_{e_j}$ are interior points of $e_i$ and $e_j$, respectively, and  $a_{ij}\cap\Gamma=\{s_{e_i},s_{e_j}\}$. The arc $s_i$ ($s_j$) is the segment of $e_i$ ($e_j$) from $v$ to $s_{e_i}$ ($s_{e_j}$), while $s_{i}$, $s_{j}$ and $a_{ij}$ generate a triangle $\alpha_{ij} := s_{i} \circ a_{ij} \circ s_j^{-1}$. 

Three (non-planar) links define a tetrahedra (see Fig. \ref{tetrahedron}).
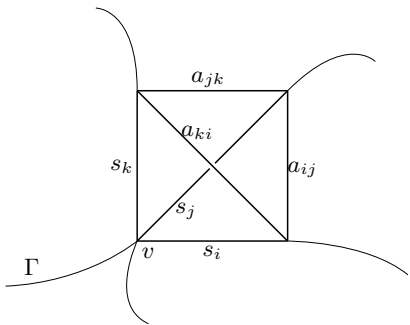
\begin{figure}
  \begin{center}$
\begin{array}{c}
	\ifx\JPicScale\undefined\def\JPicScale{0.5}\fi
\psset{unit=\JPicScale mm}
\psset{linewidth=0.3,dotsep=1,hatchwidth=0.3,hatchsep=1.5,shadowsize=1,dimen=middle}
\psset{dotsize=0.7 2.5,dotscale=1 1,fillcolor=black}
\psset{arrowsize=1 2,arrowlength=1,arrowinset=0.25,tbarsize=0.7 5,bracketlength=0.15,rbracketlength=0.15}
\begin{pspicture}(0,0)(109,84)
\psline[linewidth=0.4](37,22)(77,62)
\psline[linewidth=0.4,border=0.8](37,62)(77,22)
\psline[linewidth=0.4](37,22)(37,62)
\psline[linewidth=0.4](37,22)(77,22)
\psline[linewidth=0.4](77,62)(77,22)
\psline[linewidth=0.4](37,62)(77,62)
\pscustom[linewidth=0.2]{\psbezier(37,62)(37,84)(26,84)(26,84)
\psbezier(26,84)(26,84)(26,84)
}
\pscustom[linewidth=0.2]{\psbezier(77,62)(92.56,77.56)(100.33,69.78)(100.33,69.78)
\psbezier(100.33,69.78)(100.33,69.78)(100.33,69.78)
}
\pscustom[linewidth=0.2]{\psbezier(109,13)(98,22)(77,22)(77,22)
\psbezier(77,22)(77,22)(77,22)
}
\pscustom[linewidth=0.2]{\psbezier(2,10)(24,11)(37,22)(37,22)
\psbezier(37,22)(37,22)(37,22)
}
\pscustom[linewidth=0.2]{\psbezier(40,0)(29,5)(37,22)(37,22)
\psbezier(37,22)(37,22)(37,22)
}
\rput(40,19){$v$}
\rput(57,19){$s_i$}
\rput(50,30){$s_j$}
\rput(33,42){$s_k$}
\rput(81,41){$a_{ij}$}
\rput(56,65){$a_{jk}$}
\rput(53,51){$a_{ki}$}
\rput(9,15){$\Gamma$}
\end{pspicture}
\end{array}
$
    \parbox{9cm}{\caption[]{\label{tetrahedron} \small
       An elementary tetrahedron $\Delta \in T$ constructed
      by adapting it to a graph $\Gamma$ which
       underlies a cylindrical function.}}
  \end{center}
\end{figure}
 The full triangulation $T$ contains the tetrahedra obtained by considering all the incident links at a given node and all the possible nodes of the graph $\Gamma$. Now we can decompose \eqref{H_E_N0} into the sum of the following term per each tetrahedra $\Delta$ of the triangulation $T$  
\begin{eqnarray}
  \label{Ham_T10}
  H[N] = \sum_{ \Delta \in T}
        {-2}\, \int_{\Delta} d^3x \, N 
            \ \epsilon^{abc}\ {\rm Tr}(F_{ab}
            \{ A_c, V \}) ~.
          \label{Ham_T20}
\end{eqnarray}

 The connection $A$ and the curvature $F$ are regularized by writing them in terms of holonomy $h^{(m)}_{s}:=h[s]\in\group{SU}{2}$ in a general representation $m$ along the segments $s_i$ and the loop $\alpha_{ij}$, respectively. This yields
 \be
   \label{Hm_delta:classical10}
   H^m_{\Delta}[N]:= \frac{N(\vgraph)}{N^2_m} \,  \, \epsilon^{ijk} \,
   \mathrm{Tr}\Big[h^{(m)}_{\alpha_{ij}} h^{(m)-1}_{s_{k}} \big\{h^{(m)}_{s_{k}},V\big\}\Big] ~,
 \ee
 the trace being in an arbitrary irreducible representation $m$: $\mathrm{Tr}_{m}[D]=\mathrm{Tr}[D^{(m)}(U)]$ where $D^{(m)}$ is a matrix representation of $U\in\group{SU}{2}$, while $N_m^2=\mathrm{Tr}_m[\tau^i \tau^i]=-(2m+1)m(m+1)$ and $h^{(m)}=D^{(m)}(h)$.  As shown in \cite{Gaul:2000ba}, the right-hand side of Eq. (\ref{Hm_delta:classical10}) converges to the Hamiltonian constraint \eqref{Ham_T20} if the triangulation is sufficiently fine. The expression (\ref{Hm_delta:classical10}) can finally  be promoted to a quantum operator, since the volume and the holonomies have corresponding well-defined operators in $\mathcal{H}^{kin}$ and replacing the Poisson brackets with the commutator $\{,\}\rightarrow -\frac{i}{\hbar}[,]$ we get:
 \be
   \label{Hm_delta:quantum}
   \hat{H}^m_{\Delta}[N]:= N(\vgraph)C(m) \,  \, \epsilon^{ijk} \,
   \mathrm{Tr}\Big[\hat{h}^{(m)}_{\alpha_{ij}} \hat{h}^{(m)-1}_{s_{k}} \big[\hat{h}^{(m)}_{s_{k}},\hat{V}\big]\Big] .
 \ee
 where $C(m)=\frac{-i}{8\pi \gamma l^2_p N^2_m}$.
  The lattice spacing $\epsilon$ of the triangulation $T$ acts as a regularization parameter and it can be removed in a suitable operator topology in the space of s-knots, see  \cite{qsd} for details. This is essentially due to the fact that via a diffeomorphisms it is possible to change $\epsilon$, thus the result of the computation of $H^m_{\Delta}$ over diffeomorphisms-invariant states does not depend on such a regulator.    
 \end{itemize}
 
Remarkably it's possible to formally write solutions to the quantum Hamiltonian constraint: these are linear combinations of spinnetworks based on graph with ``dressed'' nodes (see \cite{thiemann book}) characterized by ``extraordinary links'', {\it i.e.} links with three-valent nodes as boundary attached to two collinear links. Due to the particular nature of the ``dressed'' spinnetworks the procedure described gives an anomaly free quantization of the Dirac algebra. However these solutions are only formal because the explicit expression of the matrix elements of $\hat{H}$ is very complicated \cite{io e antonia} and it's unknown in a closed form  because of the presence of the volume operator  (for which only numerical calculations are available for arbitrary valence and spins \cite{Brun}).
In the quantum-reduced model that we are going to introduce, instead the volume operator is diagonal and this will allow us to explicitly compute the matrix element of $\hat{H}$, opening the way to construct the physical quantum states.

\section{Bianchi models} \label{Bianchiom}
The early phases of the Universe are described by skipping the assumptions of the FRW model, {\it i.e.} isotropy and homogeneity. The relaxing of the former leads to the Bianchi models for the Universe (see \cite{mbb} for a recent review), which are described by the following line element 
\begin{equation} 
ds^2 = N^2(t)dt^2 - e^{2\alpha(t)}(e^{2\beta(t)})_{ij}\,\omega^i \otimes \omega^j\ ,\label{Bel}
\end{equation}

$\alpha$, $N$ and $\beta_{ab}$ depending on time coordinates. $\alpha$ determines the total volume, while the matrix $\beta_{ab}$ describes local anisotropies and it can be taken as diagonal and with a vanishing trace, such that two independent components remain. The fiducial 1-forms $\omega^i$ determine the fiducial metric on the spatial manifold. 

For a Bianchi model, the homogeneity of the fiducial metric allows to define some structure constant $C^i_{jk}$ as follows
\be
d\omega^i=C^i_{jk}\omega^j\wedge\omega^k.
\ee

Each model is determined by $C^i_{jk}$ and the Bianchi type I, II and IX are characterized by $C^i_{jk}=\{0,\delta^i_1\epsilon^1_{jk},\epsilon^i_{\phantom1jk}\}$, respectively. In the following, we will restrict our attention to the so-called class A models for which $C^i_{ij}=0$.

Densitized 3-bein vectors can be determined from the expression of the spatial metric tensor in Eq. (\ref{Bel}). However, it is not possible to fix uniquely $E^a_i$ because one is always free to perform a rotation in the internal space which does not modify the metric tensor. A useful choice is to set $E^a_i$ parallel to the vectors $\omega_i$, defined as $\omega^i(\omega_j)=\delta^i_j$, such that it is possible to separate gauge and dynamical degrees of freedom \cite{boj00}. It is worth noting how this choice implies a gauge fixing of the symmetry under internal rotations. The associated gauge-fixing condition reads \cite{cm12,cmm12} 
\be
\chi_i=\epsilon_{ij}^{\phantom1k}E_k^a\omega_a^j.\label{chicl}
\ee

At the end, the following expression for densitized 3-bein vectors is inferred
\begin{equation}
E^a_i=p^i(t) \omega \omega^a_i,\qquad p^i=e^{2\alpha}e^{-\beta_{ii}},\label{E}
\end{equation}

$\omega$ being the determinant of $\omega^j_b$, while {\it the index $i$ is not summed}. In the following, repeated gauge indexes will not be summed while the Einstein convention will still be applied to the indexes in the tangent space. The associated Ashtekar-Barbero-Immirzi connections can be inferred by evaluating the extrinsic curvature $K_{ab}$ and the 3-dimensional spin connections $\omega_{ija}$. The extrinsic curvature involves time-derivatives of the 3-metric and $K_a^i=K_{ab}e^{ib}$ reads
\be
K^i_{a}=\frac{1}{2N}\partial_th_{ab}=\frac{1}{2N}(\dot{\alpha}+\dot{\beta_{ii}})e^\alpha e^{\beta_{ii}}\omega^i_a,\label{k}
\ee

while the expression of the spin connection $\omega_{ija}$ is given by
\be
\omega_{ija}=\frac{1}{2}a_k^{-1}(a_ia_j^{-1}a_k^{-1}C^i_{jk}+a_ja_k^{-1}a_i^{-1}C^j_{ki}-a_ka_i^{-1}a_j^{-1}C^k_{ij}),
\label{omega}\ee

where $a_i=e^{\alpha+\beta_{ii}}$.

The connection $A^i_a$ is given by the sum of $\gamma K_a^i$ and $\frac{1}{2}\epsilon^{ijl}\omega_{jla}$, and it can be written as  
\begin{equation}
A^i_a= c_i(t)\omega^i_a,\qquad c_i=\left(\frac{\gamma}{N}(\dot{\alpha}+\dot{\beta_{ii}})+\alpha_i\right)e^\alpha e^{\beta_{ii}},\label{A}
\end{equation}

where $\alpha_i$ depends on the kind of Bianchi model adopted ($\frac{1}{2}\sum_{j,k}\epsilon^{ijk}\omega_{jka}=\alpha_i\omega^i_a$). 

\subsection{Loop Quantum Cosmology}

The LQC formulation of homogeneous Bianchi models implements the quantization procedure in the reduced phase space parametrized by $\{c_i,p^j\}$ \cite{ABL03}. 

The induced symplectic structure leads to the following Poisson brackets
\be
\{p^i(t),c_j(t)\}_{PP}=\frac{8\pi G}{V_0}\gamma \delta^i_j,\label{poisson}
\ee

the other vanishing, where $V_0$ denotes the volume of the fiducial cell on which the spatial integration occurs.

The Hilbert space is defined by addressing a polymer-like quantization and it turns out to be the direct product of three Bohr compactifications of the real line, $\mathcal{H}=\mathcal{L}^2(\textbf{R}^3_{Bohr},d\vec\mu)$, one for each fiducial direction. A generic basis element is thus the direct product of three quasi-periodic functions, {\it i.e.}
\be 
\psi_{\vec{\mu}}(c_1,c_2,c_3)=\otimes_{i} e^{i\mu_ic_i},
\ee

$\vec{\mu}=\{\mu_i\}$ being real numbers. The operators associated with momenta $p^i$ act as follows
\be 
\hat{p}^i\psi_{\vec{\mu}}(c_1,c_2,c_3)=8\pi\gamma l_P^2 \mu_i\psi_{\vec{\mu}}(c_1,c_2,c_3).
\ee

The scalar constraint is derived by rewriting the one of LQG (\ref{Hm_delta:classical10}) in terms of the holonomies associated with the connections (\ref{A}) and of the reduced volume operator $V=V_0 \sqrt{p^1p^2p^3}$. However, the area of the additional plaquette $\alpha_{ij}$ cannot be sent to $0$. The difference with respect to the full theory can be traced back to the loss of diffeomorphisms symmetry, which was responsible for the restriction to s-knots. This issue has been solved by evaluating the scalar constraint at some fixed non-vanishing values $\bar\mu_i\bar\mu_j$ for the area of the plaquettes $\alpha_{ij}$. These values are related with the scale at which the discretization of the geometry in LQG occurs \cite{qnb}. 
The resulting dynamics has been analyzed for Bianchi I, II and IX models \cite{qnb,bohe,ashedI,mm,ashedII} and the presence of $\bar\mu$'s provides a nontrivial evolution for the early phase of the Universe, whose most impressive consequence is the replacement of the initial singularity with a bounce.    

Therefore, in LQC the parameters $\bar\mu$'s contain all the information on the quantum geometry underlying the continuous spatial picture and, at the same time, they are responsible for the departure from the standard Big Bang paradigm. 

However this construction only mimics the original LQG quantization and even if  it is well defined on physical ground there is still a gap between the full theory and this scheme. The formalism that we are going to introduce is instead obtained by a direct reduction from the full theory at a quantum level and it could shed light on the $\bar{\mu}$ scheme at the base of LQC.

\section{Inhomogeneous variables} \label{BianchiI}

Our aim is to consider a weaker classical reduction of the full phase-space with respect to the one used in LQC, in such a way that a reduced diffeomorphisms invariance is retained and there is then more freedom in the regularization of the superhamiltonian operator. In this respect, we will consider an inhomogeneous extension of the Bianchi I model.

The Bianchi I model describes a spatial manifold isomorphic to a 3-dimensional hyperplane. The structure constants $C^i_{jk}$ vanish and the 1-forms $\omega^i$ can be taken as $\omega^i=\delta^i_adx^a$. The metric of Bianchi I model can be written in Cartesian coordinates as follows
\be
ds^2_I=N^2dt^2-a_1^2(t)dx^1\otimes dx^1-a_2^2(t)dx^2\otimes dx^2-a_3^2(t)dx^3\otimes dx^3,\label{BI}
\ee

$a_i$ $(i=1,2,3)$ being the three scale factors depending on the time variable only. 

Let us now consider the following inhomogeneous extension of the line element (\ref{BI})
\be
ds^2_I=N^2(x,t)dt^2-a_1^2(t,x)dx^1\otimes dx^1-a_2^2(t,x)dx^2\otimes dx^2-a_3^2(t,x)dx^3\otimes dx^3,
\ee

in which each scale factor $a_i$ is a function of time and of the spatial coordinates. As soon as the gauge condition (\ref{chicl}) holds the densitized inverse 3-bein vectors read
\be
E^a_i=p^i(t,x)\delta^a_i,\qquad p^i=\frac{a_1a_2a_3}{a_i},\label{Ein}
\ee

{\it i.e.} they take the same expression as in the relation (\ref{E}), the only difference being that now reduced variables $p^i$ depend also on spatial coordinates. A similar result is obtained for the projected extrinsic curvature, {\it i.e.}
\be
K^i_a=\frac{1}{N}\dot{a}_i(t,x) \delta^i_a,
\ee  

while the spin connections $\omega_{ija}$ for the inhomogeneous model are given by
\be
\omega_{ija}=a_i^{-2}a_j^{-1}\delta_a^i\delta_j^b\partial_b a_{i}-a_j^{-2}a_i^{-1}\delta_a^j\delta_i^b\partial_b a_{j}.
\label{omegain}\ee

 At this point let us consider two different cases: 1) the reparametrized Bianchi I model and 2) the generalized Kasner solution within a fixed Kasner epoch. 

In a reparametrized Bianchi I model we assume that each scale factor is a function of time and of the corresponding Cartesian coordinate $x^i$ only, {\it i.e.}  
\be
a_i=a_i(t,x^i),
\ee

such that $\partial_b a_i\propto \delta_b^i$ and the spin connections $\omega_{ija}$ vanish identically. Obviously, the dependence on $x^i$ is fictitious and it can always be avoided by a diffeomorphisms, so finding the homogeneous Bianchi I model. However, the reparametrized model is endowed with an additional gauge symmetry,  which will have a key-role in the development of the quantum theory.  

The same result concerning the vanishing of spin connections can also be obtained in the limit in which the spatial gradients of the metric components can be neglected with respect to the time derivatives. This approximation scheme corresponds to the notion of ``local homogeneity'', which is implemented when the BKL mechanism is extended to the generic cosmological solution \cite{BKL,mbb}. This is done by considering the generalized Kasner model \cite{gkas}, which describes the behavior of the generic cosmological solution during each Kasner epoch. This model has been realized by considering an extension of the Kasner solution, in which the Kasner exponents are functions of spatial coordinates. Indeed, in general the fiducial vectors do not coincide with the ones of the homogeneous Bianchi I model and they are subjected to a rotation signaling the transition to a new epoch. Nevertheless, within each epoch, one can neglect the rotation of Kasner axes and take at the leading order the fiducial vectors $\omega^i_a=\delta^i_a$.

Therefore, in both cases 1) and  2) the connections retain the same expression as in the homogeneous case, but reduced variables depend on spatial coordinates as follows
\be
A^i_a(t,x)= c_i(t,x)\delta^i_a,\qquad c_i(t,x)=\frac{\gamma}{N}\dot{a_i}.\label{Ain}
\ee  


The Poisson brackets between $A^i_a$ and $E^a_i$ induce the following Poisson algebra
\be
\{p^i(x,t),c_j(y,t)\}=8\pi G\gamma \delta^i_j\delta^{3}(x-y),\label{poissonin}
\ee

the other vanishing.

Since we did not impose homogeneity, the SU(2) Gauss constraint $G_i$ and the super-momentum constraint $H_a$ do not vanish identically. In particular, $G_i$ reads
\be
G_i=\delta_i^a\partial_ap^i=\partial_ip^i,\label{gauss}
\ee 

while the generator of 3-diffeomorphisms takes the following expression
\begin{eqnarray}
D[\vec\xi]=\int \xi^a[H_a-A^i_aG_i]d^3x=\sum_i\int [\xi^ap^i\partial_ac_i+(\partial_i \xi^i) p^ic_i] d^3x,\label{rdiff}
\end{eqnarray} 

$\xi^a$ being arbitrary parameters, while $\xi^i=\xi^a\delta_a^i$.

\section{Reduced quantization for inhomogeneous Bianchi model}\label{reduced}

Let us now discuss how the quantization of inhomogeneous Bianchi models can be performed in reduced phase space. 

In this case, one should define the Hilbert space for functionals of reduced variables $c_i$, whose conjugate variables are $p^i$, and consider the set of reduced constraints. In particular, the SU(2) Gauss constraint is replaced by the conditions (\ref{gauss}), which for a given $i$ can be regarded as a $U(1)$ Gauss constraint along the one-dimensional space generated by the vector dual to $\omega^i=\delta^i_a dx^a$ \emph{i.e} $\partial_i=\delta^a_i\partial_a$. We denote the $U(1)$ group of transformations generated by $G_i$ as $U(1)_i$. Since $\{G_i,G_j\}=0$, the $U(1)_i$ transformations are all independent from each other.

A convenient choice of variables  for the loop quantization is to consider the $U(1)_i$ holonomies for the connections $c_i$ along the edges $e_i$ parallel to $\partial_i$, {\it i.e.}
\be
{}^{\mathrm{red}}h_{e_i}=P(e^{i\int_{e_i}c_idx^i}).
\label{hred}
\ee

Hence, we are not dealing with a $U(1)^3$ gauge theory on a 3-dimensional space, since holonomies associated with different $U(1)_i$ have support on different edges $e_i$. What we have is the direct product of three 1-dimensional $U(1)$ gauge theories. 
 
The Hilbert space can be labeled by \emph {reduced graphs} $\Gamma$, which are cuboidal lattices made by the union of (at most) 6-valent vertices with the ingoing and outgoing edges of the kind $e_i$, and it can be defined as the direct product of the space of square integrable functionals over the $U(1)_i$ group elements associated with each $e_i$, {\it i.e.}
\be
^{red}\mathcal{H}=\bigotimes_{i=1}^3\bigotimes_{e_i\in\Gamma}\mathcal{L}^2\left(U(1)_i,d\mu_i\right), 
\ee
$d\mu_i$ being the $U(1)_i$ Haar measure.

A generic element is given by taking the direct product of $U(1)_i$ networks over $e_i$ and they read
\be
\psi_{\Gamma}=\bigotimes_{i=1}^3\bigotimes_{e_i\in \Gamma}\psi_{e_i},
\ee

 where  $\psi_{e_i}$ is a $U_i(1)$  function, which can be expanded in $U(1)_i$ irreducible representations as follows
\be
\psi_{e_i}=\sum_{n_i}e^{in_i\theta^i}\psi^{n_i}_{e_i}, \label{uired}
\ee
$\theta^i$ being the parameter over the $U(1)_i$ group, while $n_i$ denotes the $U(1)_i$ quantum number.

Momenta $p^i$ have to be smeared over the surfaces $S^i$ dual to $e_i$ and the associated operators can be inferred by quantizing the Poisson algebra (\ref{poisson}), so finding

\be
\hat{p}^l(S^i)\psi_{e_i}=8\pi\gamma l_p^2\delta^l_i \sum_{n_i}n_ie^{in_i\theta^i}\psi^{n_i}_{e_i}.\label{redei}
\ee

In order to develop the gauge-invariant Hilbert space $^{red}\mathcal{H}^{G_i}$ in which the conditions (\ref{gauss}) are solved, one must insert the invariant intertwiners associated with the three $U(1)_i$ groups. These intertwiners map $U(1)_i$ group elements each other for a fixed value of $i$. This means that they do not provide us with a real 3-dimensional vertex structure, since they connect only group elements defined over intersecting edges parallel to the same vector field $\partial_i$.

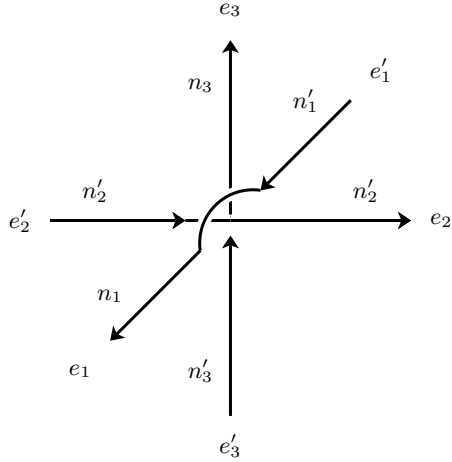
\begin{figure}
$
\ifx\JPicScale\undefined\def\JPicScale{2}\fi
\psset{unit=\JPicScale mm}
\psset{linewidth=0.3,dotsep=1,hatchwidth=0.3,hatchsep=1.5,shadowsize=1,dimen=middle}
\psset{dotsize=0.7 2.5,dotscale=1 1,fillcolor=black}
\psset{arrowsize=1 2,arrowlength=1,arrowinset=0.25,tbarsize=0.7 5,bracketlength=0.15,rbracketlength=0.15}
\begin{pspicture}(0,0)(32,29)
\psline[linewidth=0.2,arrowsize=0.75 2,arrowlength=0.75]{<-}(18,27)(18,15)
\psline[linewidth=0.2,border=0.3](15,15)(18,15)
\rput{0}(19.5,13.5){\psellipticarc[linewidth=0.2,border=0.3](0,0)(3.54,3.54){81.87}{188.13}}
\psline[linewidth=0.2,arrowsize=0.75 2,arrowlength=0.75]{->}(26,23)(20,17)
\psline[linewidth=0.2,arrowsize=0.75 2,arrowlength=0.75]{->}(16,13)(10,7)
\rput(32,15){$e_2$}
\rput(26,21){}
\rput(8,5){$e_1$}
\rput(18,29){$e_3$}
\rput(28,25){$e'_1$}
\rput(4,15){$e'_2$}
\rput(18,0){$e'_3$}
\psline[linewidth=0.2,arrowsize=0.75 2,arrowlength=0.75]{<-}(18,14)(18,2)
\psline[linewidth=0.2,border=0.3,arrowsize=0.75 2,arrowlength=0.75]{->}(18,15)(30,15)
\psline[linewidth=0.2,border=0.3,arrowsize=0.75 2,arrowlength=0.75]{->}(6,15)(15,15)
\rput(9,17){$n'_2$}
\rput(27,17){$n'_2$}
\rput(10,10){$n_1$}
\rput(23,23){$n'_1$}
\rput(16,24){$n_3$}
\rput(16,5){$n'_3$}
\end{pspicture}\nonumber$
\caption{The representation of a vertex in reduced quantization: the quantum numbers $n_1,n_2,n_3$ are conserved along the directions $i=1,2,3$, respectively.} \label{vertrq}
\end{figure}

At a single vertex $v$ one can have at most two $U(1)_i$ group elements for a given $i$: the ones associated with the two edges $e_i$ and $e'_i$ emanating form $v$ (see Figure \ref{vertrq}). 

As soon as $\psi_{e_i}$ and $\psi_{e'_i}$ are expanded in irreducible representations $\psi^{n_i}_{e_i}$ and $\psi^{n'_i}_{e'_i}$ (\ref{uired}), respectively, the invariant intertwiner selects those representations for which $n_i=n'_i$.

Therefore, the projection to $^{red}\mathcal{H}^{G_i}$ implies that the $U(1)_i$ quantum numbers are preserved along each integral curve of the vectors $\partial_i$.  


\subsection{Diffeomorphisms}\label{sdiff}

The conditions (\ref{Ein}) and (\ref{Ain}) imply a partial gauge fixing of the diffeomorphisms symmetry. In fact, under a generic 3-diffeomorphisms connections and momenta transform as follows
\be
\delta_\xi A^i_a=\xi^b\partial_bA^i_a+\partial_a\xi^bA^i_b,\qquad \delta_\xi E^a_i=\xi^b\partial_b E^a_i-\partial_b\xi^aE^b_i.
\ee

Starting from the expression (\ref{A}), one gets
\be
\delta_\xi A^i_a=\xi^b\partial_bc_i\delta^i_a+\xi^bc_i\partial_b\delta_a^i+\partial_a\xi^b\delta^i_bc_i=
\xi^b\partial_bc_i\delta^i_a+\partial_a\xi^ic_i.
\ee 

It is worth noting that for arbitrary $\xi^a$ the connection cannot be written as in (\ref{A}). This feature signals that by choosing connections as in (\ref{Ain}) we are actually performing a partial gauge-fixing of the diffeomorphisms group. The same result is obtained for $E^a_i$. However, there is a residual set of admissible transformations which preserve the conditions (\ref{A}) and (\ref{E}) and they are those ones for which
\be
\partial_a\xi^i\propto \delta^i_a\rightarrow \xi^i=\xi^i(x^i).\label{reddiff}
\ee   

As soon as the condition above holds, each $\xi^i$ is the infinitesimal parameter of an arbitrary translation along the direction $i$ and a rigid translation along other directions. We denote this transformations as \emph{reduced diffeomorphisms} $\tilde{\varphi}_\xi$. We are going to show how the constraint (\ref{rdiff}) implies the invariance under reduced diffeomorphisms.

In reduced phase-space, the constraint (\ref{rdiff}) acts on a reduced holonomy (\ref{hred}) as follows
\be
\hat{D}[\vec{\xi}]{}^{\mathrm{red}}h_{e_i}=8\pi\gamma l^2_P\int_{e_i} {}^{\mathrm{red}}h_{e_i(0,s')} (\xi^b\partial_bc_i+(\partial_i \xi^i)c_i){}^{\mathrm{red}}h_{e_i(s',1)}dx^i(s'),  \label{dt}  
\ee 

$e_i(0,s')$ and $e_i(s',1)$ being the edges from $s=0$ to $s=s'$ and from $s=s'$ to $s=1$, respectively. 

The transformation (\ref{dt}) has to be compared with the changing induced by a reduced diffeomorphisms $\tilde\varphi_\xi:x^a(s)\rightarrow x'^a(s)=x^a(s)+\xi^a$ under the condition (\ref{reddiff}). A diffeomorphisms $\varphi$ maps an edge $e_i$ into one which is generically not of the reduced class. In fact the tangent vector at the leading order is given by the following expression
\begin{eqnarray}
\frac{dx'^a}{ds}=\frac{dx^a}{ds}+\partial_b\xi^a\frac{dx^b}{ds}\propto \delta^a_i+
\partial_b\xi^a\delta^b_i=\delta^a_i+\partial_b\xi^j\delta^a_j\delta^b_i.
\end{eqnarray}
The second term on the right side gets contributions also from the fiducial vectors $\partial_j$ with $j\neq i$, such that the tangent vector of $\varphi(e_i)$ is not proportional to $\partial_i$. However, if one considers the reduced class of transformations (\ref{reddiff}), these additional contributions vanish and the tangent vector of $\tilde\varphi(e_i)$ is parallel to $\partial_i$. Hence, reduced diffeomorphisms $\tilde\varphi$ map reduced edges $e_i$ each others.

The holonomy along $\tilde\varphi_\xi$ is thus given by 
\begin{equation}
h_{\tilde\varphi_\xi(e_i)}=P(e^{\int c_i(x')\delta^i_adx'^a}),
\end{equation}
and by computing the integrand one gets
\begin{eqnarray}
c_i(x')\delta^i_adx'^a=c_i(x)\delta^i_adx^a+\xi^b\partial_bc_i\delta^i_adx^a+c_i(x)\partial_a \xi^i dx^a.
\end{eqnarray}

From the expression above and by considering that $dx^a=\delta^a_i dx^i(s)$ the following relation follows   
\begin{eqnarray}
h_{\tilde\varphi_\xi(e_i)}-h_{e_i}=P(e^{\int c_i(x')\delta^i_adx'^a})-P(e^{\int c_i(x)\delta^i_adx^a})=\int_{e_i} {}^{\mathrm{red}}h_{e_i(0,s')} (\xi^b\partial_bc_i+(\partial_i \xi^i)c_i){}^{\mathrm{red}}h_{e_i(s',1)}dx^i(s').
\end{eqnarray}
which coincides with the expression (\ref{dt}). Therefore, the reduced diffeomorphisms 
$\tilde\varphi$ (\ref{reddiff}) map reduced holonomies into reduced holonomies and they are associated with the action of the relic diffeomorphisms constraint (\ref{rdiff}) in reduced phase-space. This residual symmetry can be used to define  reduced knot classes as in the full theory. 

\subsection{Dynamics}

\begin{figure}
$
\ifx\JPicScale\undefined\def\JPicScale{1}\fi
\unitlength \JPicScale mm
\begin{picture}(115,70)(0,0)
\linethickness{0.3mm}
\put(10,50){\line(1,0){20}}
\put(30,50){\vector(1,0){0.12}}
\linethickness{0.3mm}
\put(30,50){\line(1,0){20}}
\put(50,50){\vector(1,0){0.12}}
\linethickness{0.3mm}
\put(30,30){\line(0,1){20}}
\put(30,50){\vector(0,1){0.12}}
\linethickness{0.3mm}
\put(30,50){\line(0,1){20}}
\put(30,70){\vector(0,1){0.12}}
\linethickness{0.3mm}
\put(90,30){\line(0,1){20}}
\put(90,50){\vector(0,1){0.12}}
\linethickness{0.3mm}
\put(90,50){\line(0,1){20}}
\put(90,70){\vector(0,1){0.12}}
\linethickness{0.3mm}
\put(70,50){\line(1,0){20}}
\put(90,50){\vector(1,0){0.12}}
\linethickness{0.3mm}
\put(90,50){\line(1,0){20}}
\put(110,50){\vector(1,0){0.12}}
\linethickness{0.3mm}
\put(35,45){\line(1,0){10}}
\put(45,45){\vector(1,0){0.12}}
\put(35,35){\line(0,1){10}}
\put(45,35){\line(0,1){10}}
\put(35,35){\line(1,0){10}}
\linethickness{0.3mm}
\put(100,40){\line(0,1){10}}
\linethickness{0.3mm}
\put(90,40){\line(1,0){10}}
\put(10,55){\makebox(0,0)[cc]{$n$}}

\put(50,55){\makebox(0,0)[cc]{$n$}}

\put(50,40){\makebox(0,0)[cc]{$1$}}

\put(25,30){\makebox(0,0)[cc]{$p$}}

\put(25,65){\makebox(0,0)[cc]{$p$}}

\put(70,55){\makebox(0,0)[cc]{$n$}}

\put(110,55){\makebox(0,0)[cc]{$n$}}

\put(100,55){\makebox(0,0)[cc]{$n+1$}}

\put(105,45){\makebox(0,0)[cc]{$1$}}

\put(85,30){\makebox(0,0)[cc]{$p$}}

\put(85,40){\makebox(0,0)[cc]{$p+1$}}

\put(85,65){\makebox(0,0)[cc]{$p$}}

\end{picture}
$
\caption{The action of the operator associated with the curvature changes $U(1)_i$ quantum numbers such that it maps the state out of the gauge-invariant Hilbert space (we did not drawn the edges along the third direction).}\label{dinrq}
\end{figure}
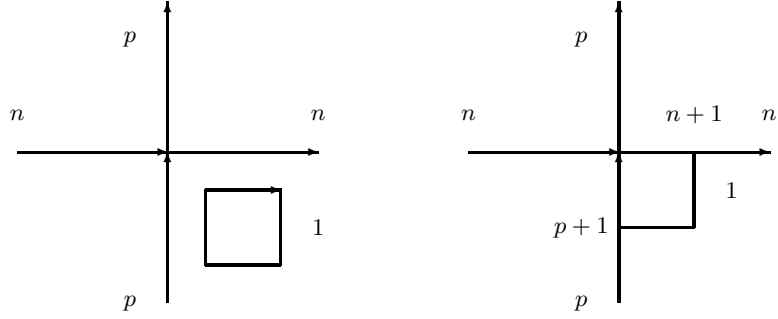

The superHamiltonian operator in reduced-phase space takes the following form
\be
H[N]=\int d^3x N\left[\sqrt{\frac{p^1p^2}{p^3}}c_1c_2+\sqrt{\frac{p^2p^3}{p^1}}c_2c_3+\sqrt{\frac{p^3p^1}{p^2}}c_3c_1\right], 
\ee

and the quantization of this expression requires to i)give a meaning to the operator $1/\sqrt{p^i}$, ii) replace $c_i$ with some expression containing holonomies. These are the standard issues one encounters in LQG, which are solved by quantizing the expression (\ref{Hm_delta:classical10}). Therefore, the quantization of the superHamiltonian operator in the reduced model can be realized by implementing in the reduced Hilbert space the procedure adopted in the full theory. This can be done formally by replacing $SU(2)$ group elements with $U(1)_i$ ones and by defining a cubulation of the spatial manifold, such that the loop $\alpha_{ij}$ is a rectangle with edges along fiducial vectors.  Unfortunately, the resulting expression for the superHamiltonian operator regularized \`a la Thiemann  is not defined in $^{red}\mathcal{H}^{G_i}$. This is due to the fact that the operator $h_{\alpha_{ij}}$ increases (decreases) the $U(1)_i$ ($U(1)_j$) quantum number associated with the segment $s_i$ ($s_j$). As a consequence, the $U(1)_i$ quantum number is not conserved along the edge $e_i$ and the $U(1)_i$ symmetry is broken (see figure \ref{dinrq}).

Therefore, it cannot be given a proper definition of the superHamiltonian operator in reduced quantization. This is due to the lack of a real 3-dimensional vertex structure, which   instead can be inferred starting from the full LQG theory.

\section{Cosmological LQG} \label{CLQG}
Let us now discuss how to realize in the SU(2) kinematical Hilbert space of LQG  $\mathcal{H}^{kin}$ the conditions (\ref{Ein}) and (\ref{Ain}) via a reduction from SU(2) to U(1) group elements. 

At first, we impose the restriction to edges $e_i$ parallel to fiducial vectors $\partial_i$ and we discuss the fate of diffeomorphisms invariance. Then, we will deal with the restriction from $SU(2)$ to $U(1)$ group elements and with the relic features of the original $SU(2)$ invariance. 

\subsection{Quantum Diff-Constraint}

The restriction to cylindrical functionals over edges $e_i$ implies the kind of restriction on the diffeomorphisms transformations which we discussed in section \ref{sdiff}. We can implement this feature on a quantum level via the action of a projector $P$ onto the space $\mathcal{H}_P$ made of holonomies along reduced graphs (edges $e_i$ adapted to the $\omega_i$). This projector $P$ acting on ${\mathcal H}^{kin}$ is then nonvanishing only for holonomies along edges $e_i$.

Let us consider a generic diffeomorphisms $\varphi_\xi$, whose associated operator $U(\varphi_\xi)$ in the space of cylindrical functional acts on a generic holonomy $h_e$ along an edge $e$ as follows
\be
\hat{U}(\varphi_\xi)h_e=h_{\varphi_\xi(e)}.
\ee

The projection of $U(\varphi_\xi)$ in the graph-reduced Hilbert space $\mathcal{H}_P$ is given by
\be
{}^{red}\!\hat{U}(\varphi_\xi)=P\hat{U}(\varphi_\xi)P,
\ee

where $Ph_e=h_e$ if $e=e_i$ for some $i$, otherwise it vanishes. The action of ${}^{red}\!U(\varphi)$ on a graph-reduced holonomy $h_{e_i}$ reads then
\be
{}^{red}\!\hat{U}(\varphi)h_{e_i}=P\hat{U}(\varphi_\xi)Ph_{e_i}=P\hat{U}(\varphi_\xi)h_{e_i}=P
h_{\varphi_\xi(e_i)}.\label{redu}
\ee

As we pointed out in section \ref{sdiff}, $\varphi_\xi(e_i)$ is parallel to $\omega_i$ if $\varphi$ is a reduced diffeomorphisms $\tilde{\varphi}$. Hence, the relation (\ref{redu}) is nonvanishing only if $\varphi=\tilde\varphi$ and one finds 
\be
{}^{red}\!\hat{U}(\varphi)=\hat{U}(\tilde\varphi).
\ee

Therefore, in $\mathcal{H}_P$ the relic diffeomorphisms are reduced ones. The development of knot classes with respect to reduced diffeomorphisms will allow us to regularize the expression of the super-Hamiltonian operator \`a la Thiemann.

\subsection{Classical holonomies and quantum reduction}

On a classical level, the $SU(2)$ holonomies ${}^Rh^j_{e_i}$ associated with connections (\ref{Ain}) are given by
\begin{equation}
{}^Rh^j_{e_i}=P(e^{i\int_{e_i}c_idx^i(s)\tau_i}), \label{rhol}
\end{equation}

$s$ being the arc length along $e_i$.

Henceforth, ${}^Rh^j_{e_i}$ are $SU(2)$ holonomies that belong to the U(1) subgroup generated by $\tau_i$ and they can be written as 
\be{}^Rh^j_{e_i}=\exp{i \alpha^i\tau_i},
\label{riduzione classica olonomia}
\ee
where $\alpha^i$ is a real number and the gauge indexes are not summed as usual.
 
These holonomies in the base $|j,m_i\rangle$ that diagonalize $\tau^i$ take the form :
\be
 <j,m_i| {}^Rh^j_{e_i}|j,n_i>=e^{i \alpha^i m_i} \delta_{m_i n_i}\label{Rhu1}
\ee
 



 Similarly, we evaluate fluxes only across the surfaces $S^i$ dual to $e_i$ and the associated fluxes in a cosmological space-time read from Eq.(\ref{E})
\be
E_i(S^k)=\int E_i^a \delta^k_a dudv=\delta_i^k\int p_idudv.
\label{riduzione classica flussi}
\ee

It is worth noting how only the diagonal components of $E_i(S^j)$ are nonvanishing. 
Now our task is to find a quantum symmetry reduction implementing consistently on the kinematical Hilbert space of LQG $\mathcal{H}^{kin}$ the classical conditions \eqref{riduzione classica olonomia} and \eqref{riduzione classica flussi}  representing the holonomized version of the variables \eqref{Ain} and \eqref{Ein}  with Poisson brackets \eqref{poissonin}. 

How can we proceed? 
 
First we observe that the skew-symmetric part of the matrix $E_i(S^j)$ can be avoided by imposing the following conditions  
\begin{equation}   
\chi_i=\sum_{l,k}\epsilon_{il}^{\phantom{12}k}E_k(S^l)=0. \label{chi}
\end{equation}

The relation above together with the SU(2) Gauss constraint constitutes a second-class system of constraints, thus it is actually a gauge-fixing. As a consequence, the condition (\ref{chi}) cannot be implemented on a SU(2) invariant quantum space according with the Dirac prescription. One possibility is to retain the full unconstrained set of configuration variables and to define the action of quantum operators starting from Dirac brackets instead of Poisson brackets. This way  however, the connections become non-commutative \cite{cm12,cmm12} and it is difficult to envisage how to carry on the quantization procedure in the full kinematical Hilbert space. 

Henceforth mimicking the procedure adopted in Spin-Foam models to impose the simplicity constraints \cite{sfsimpl}, we consider the Master constraint condition, that arises extracting the gauge invariant part of $\chi_i$,
\begin{equation}
\chi^2=\sum_i\chi_i\chi_i=\sum_{i,m,k,l}[\delta^{im}\delta_{kl}E_i(S^k)E_m(S^l)-E_i(S^k)E_k(S^i)]=0.
\label{master}
\end{equation}

By imposing the condition \eqref{master} strongly on $\mathcal{H_P}$, it will turn out that Eq. (\ref{chi}) holds weakly and the classical relation (\ref{riduzione classica flussi}) can be implemented in a proper subspace of $\mathcal{H_P}$, as soon as $p_i$ are identified with the left invariant vector fields of the $U(1)_i$ groups generated by $\tau_i$.

If $\hat{\chi}^2$ is applied to a SU(2) holonomy $h^{j}_{e_i}$ and $e_i\cap S^i=b(e_i)$, $b(e_i)$ being the beginning point of $e_i$, one finds
\begin{equation}
\hat{\chi}^2 h^{j}_{e_i}=(8\pi\gamma l_P^2)^2(\tau^2-\tau_i\tau_i) h^{j}_{e_i}\;,
\label{classic condition}
\end{equation}
thus an appropriate solution to $\chi^2=0$ is given by
\be
 \tau^k h^{j}_{e_i} =0 \;,\quad \forall k\neq i.
 \label {solution weak} 
\ee

To find the quantum states that implement $\chi^2=0$ strongly and Eq. \eqref{solution weak} weakly,  we will use projected spinnetworks \cite{alpr,dl}.

\subsection{Projected U(1)}
We now introduce the projected spin-network formalism, in which we define functions over $SU(2)$ starting from their restriction over the $U(1)_i$ subgroups generated by $\tau_i$.

This way, we lift the $U(1)_i$ group elements associated with reduced holonomies (\ref{riduzione classica olonomia}) to the $SU(2)$ elements of the full theory. This lifting will help us later in embedding reduced elements in the SU(2)-invariant Hilbert space. 
 
Let us consider the Dupuis-Livine map \cite{dl} $f: U(1)\rightarrow SU(2) $ from functions on $U(1)$ to functions on $SU(2)$ 
\be
\tilde{\psi}(g)=\int_{U(1)} dh\; K(g,h) \psi(h), \qquad g\in SU(2),
\label{projected1}
\ee
with Kernel given by 
\be
K(g,h)=\sum_{n} \int_{U(1)}  dk\; \chi^{j(n)}(gk) \chi^n(kh),
\ee
where $\chi^{j(n)}(g)$ are the $SU(2)$ characters in the $j(n)$ representations and $\chi^n(h)$ are the $U(1)$ ones,  while $j(n)$ denotes an half integer depending on an integer $n$.
 It is true that $\tilde\psi(g)|_{U(1)}=\psi$ and this implies that the image of $f$ is a subspace of the space of functions on $SU(2)$ such that 
 \be
 \tilde\psi(g)=\int_{U(1)} dh\; K(g,h) \tilde{\psi}(h), \qquad g\in SU(2),
 \ee
{\it i.e.} the function  $\tilde\psi(g)$ is entirely determined by its restriction to a $U(1)$ subgroup.
If we expand $\psi$ using the Peter-Weyl theorem we get
\be
\psi(h)=\sum_n \chi^n(h)\psi^n,
\ee
and the coefficients $\psi^n$ are given by 
\be
\psi^n=\int_{U(1)}dh \; \overline{\chi^n}(h) \psi(h) .
\ee
Eq \eqref{projected1} is then:
\be
\tilde{\psi}(g)=\sum_n \int_{U(1)} dk\;  D^{j(n)}_{m r}(g) D^{j(n)}_{rm}(k) \overline{\chi^n}(k) \psi^n,
\label{projected 2}
\ee
where $D^{j(n)}_{mr}$ are the Wigner matrices in a generic Spin base $|j,m\rangle$.
Now let us consider projected functions defined over the edge $e_i$ and let us choose the $U(1)_i$ subgroup of $SU(2)$ in the definition \eqref{projected1} as the one generated by $\tau_i$, calling its elements $k_i$ and the quantum numbers $n_i$. The previous expression becomes:

\be
\begin{split}
\tilde{\psi}(g)_{e_i}&=\sum_{n_i} \int_{U(1)_i} dk_i\;  \sum_{m,r=-j(n_i)}^{j(n_i)}{}^i\!D^{j(n_i)}_{mr}(g){}^i\!D^{j(n_i)}_{rm}(k_i) \overline{\chi^{n_i}}(k_i) \psi^{n_i}_{e_i}=\\
&=\sum_{n_i} \int_{S^1} d\theta_i\;  \sum_{mr=-j(n_i)}^{j(n_i)} {}^i\!D^{j(n_i)}_{mr}(g) e^{i m\theta_i } \delta_{mr} \overline{\chi^{n_i}}(\theta_i) \psi^{n_i}_{e_i}=\\
&=\sum_{n_i} {}^i\!D^{j(n_i)}_{m=n_i\,r=n_i
}(g) \psi^{n_i}_{e_i},
\end{split}
\label{projected 3}
\ee
${}^i\!D^{j(n_i)}_{mr}$ being the Wigner matrices in the Spin base $|j,m\rangle_i$ that diagonalize the operators $J^2$ and $J_i$ and $\theta_i$ are the coordinates on the $U(1)_i$ groups. Note that the matrices ${}^i\!D^{j}_{mr}(g)$ are obtained by the $SU(2)$ transformation $D^{j}(\vec{u}_i)$  which acts on the vector $\vec{e}_z$ sending it to the vector $\vec{u}_i = R \vec{e}_z$ as
\be
{}^i\!D^{j}_{mn}(g)={D^{j}}^{-1}_{mr}(\vec{u_i}) D^{j}_{rs}(g) {D^{j}}_{s n}(\vec{u_i}).
\label{D proiettate}
\ee

This is valid for an arbitrary degree $j(n_i)$.
Now we have to select a condition ensuring the vanishing of Eq.\eqref{classic condition} on a quantum level.
The condition on basis element of $L^2(SU(2))$ ${}^i\!D^j_{mr}(g)={}_i\!< j,m| g |j,r>_i$ reads :
\be
{}_i\!<j,m|\chi^2 g |j,r>_i={}_i\!<j,m| g |j,r>_i(j(j+1)-m^2).
\label {quantum condition}
\ee
This relation implies that if we apply $\chi^2$ to our projected spinnetworks, whose basis elements are of the form ${}^i\!D^{j(n_i)}_{n_in_i}(g)$,
by fixing $|n_i|=j(n)$ an approximate solution to $\chi^2=0$ is given as $j\rightarrow+\infty$. In the following we will consider only the plus sign \footnote{However also the case involving both signs can be considered: the basis elements of the quantum reduced Hilbert space would then be of the form ${}^i\!D^{|n_i|}_{n_in_i}(g)+{}^i\!D^{|n_i|}_{-n_i\,-n_i}(g)$}, since the opposite one can be obtained by reversing the orientation of the associated edge $e_i$. 

 It's worth noting that introducing coherent states for SU(2), defined by:
\be
|j,\vec{u}\rangle=D^j(\vec{u})|j,j\rangle=\sum_m |j,m\rangle D^j(\vec{u})_{mj}
\ee
the basis elements which are solutions of the constraint are
\be
{}^i\!D^{j}_{jj}(g)= <j,\vec{u}_i|D^j(g)|j,\vec{u}_i>\label{reduced basis elements}
\ee
for $i=1,2,3$.

Henceforth, we find

\be
\tilde{\psi}(g)_{e_i}=\sum_{j} {}^i\!D^{|j|}_{jj}(g) \psi^{j}_{e_i}.
\label{fine projected}
\ee

Basis states of this form also satisfy the condition \eqref{solution weak} weakly in fact
\be
<\tilde\psi'_i|\hat{E}_k(S^l)|\tilde\psi_i>=8\pi \gamma l_P^2 \sum_{j,j'}\psi^{j'}_{e_i}\int dg  {}^i\!D^{j'}_{j'j'}(g)\tau_k {}^i\!D^j_{jj}(g)\psi^{j}_{e_i}=0,\qquad k\neq i.\label{weak}
\ee

In this way the resulting quantum states associated with an edge $e_i$ are entirely determined by their projection into the subspace with maximum magnetic numbers along the internal direction $i$. 
We call the projected $SU(2)$ states of the form \eqref{fine projected} \emph{ quantum-reduced states} and they define a subspace of $\mathcal{H}_{P}$ that will be denoted $\mathcal{H}^{R}$. The restriction of states $\tilde\psi(g)\in \mathcal{H}^{R}$  to their $U(1)_i$ subgroup reads
\begin{equation}
\psi_{e_i}=\tilde{\psi}(g)_{e_i}|_{U(1)_i}=\sum_j e^{i \theta^i j}\psi_{e_i}^j.
\label{olonomia sulle ridotte}
\end{equation}

Therefore, the restriction to the $U(1)_i$ subgroup gives the element of  $\mathcal{H}^{red}$ (\ref{uired}).

Moreover the action of fluxes $E_l(S^k)$ on $\tilde\psi_{e_i}$ is nonvanishing only for $l=k=i$ and each $E_i(S^i)$ behaves as follows (we are assuming $S^i \cap e_i=b(e)$)
\be
\hat{E}_i(S^l)\tilde{\psi}_{e_i}=8\pi\gamma  l_P^2 \delta^l_i\sum_j j D^j_{jj}\psi^j_{e_i}.\label{flussi sulle ridotte}
\ee 

By restricting the expression (\ref{flussi sulle ridotte}) to the $U(1)$ subgroup, one gets
\be
\hat{E}_i(S^l)\tilde{\psi}_{e_i}|_{U(1)_i}=E_i(S^i)\psi_{e_i}=8\pi\gamma  l_P^2 \delta^l_i\sum_j j e^{i \theta^i j}\psi^j_{e_i}.
\ee
thus $\hat{E}_i(S^i)\psi_{e_i}$ behaves as the left-invariant vector field of the $U(1)_i$ subgroup and its action on $\psi_{e_i}$ reproduces the action of momenta in reduced quantization (\ref{redei}).
Therefore, the restriction to the $U(1)_i$ subgroup maps the quantum-reduced states, elements of $\mathcal{H}^{R}$, to the Hilbert space $\mathcal{H}^{red}$ obtained when quantizing in reduced phase-space.  

\be
\begin{CD}
\{A^i_a(x,t),E^b_k(y,t)\}\propto \delta^{i}_{k}\delta^{b}_a\delta^3(x,y) @>\mathrm{reduced\quad phase-space}>>
\{c_i(x,t),p^k(y,t)\}\propto \delta_i^k\delta^{3}(x-y)\\
@VV\mathrm{quantization}V @VV\mathrm{quantization}V\\
h_{e_i}\in SU(2),\quad \hat{E}_k(S^l)h_{e_i}\propto \delta_i^l\tau_kh_{e_i} @>\chi^2=0\quad \chi_i\sim0,>|_{U(1)_i}>\psi_{e_i}\in U(1)_i,\quad \hat{p}^k\psi^{n_i}_{e_i}\propto \delta_k^in_i\psi^{n_i}_{e_i}  
\end{CD}
\ee

Despite the possibility to project the Hilbert space of quantum reduced holonomies into the one of reduced quantization, there is a substantial difference between this two kind of reductions. The $U(1)$ representations we get are obtained by stabilizing the SU(2) group along different internal directions and the $U(1)_i$ transformations associated with different $i$ are not independent at all (they are rotations along the $i$ axis). As we will see in the next section, for this reason  some non-vanishing intertwiners exist among them. 

\subsection{Gauge invariant states}

The original SU(2) gauge invariant Hilbert space $L^2(SU(2)^L / SU(2)^N )$ is made of spinnetworks of the form \eqref{spinnet}.
These are invariant under the action of gauge transformations on the holonomies. If we define as $\hat{U}_{\mathcal{ G}}(\lambda)$ the operator that generates local $SU(2)$ gauge transformations $\lambda(x)$, it's action on basis elements of $\mathcal{H}^{kin}$ is given by
\be
\hat{U}_{\mathcal{ G}}(\lambda) D^j_{mn}(h_e)= D^j_{mn}(\lambda_{s(e)}h_e \lambda^{-1}_{t(e)}),
\ee
and by group averaging we get the projector
\be
\hat{P}_{\mathcal{ G}}= \int d\lambda \; \hat{U}_{\mathcal{ G}}(\lambda),
\label{gauge}
\ee
 acting on the source and target link  and producing the intertwiners at the nodes thanks to the formula 
 \be
\int d\lambda\;  \prod^{O}_{o=1}D^{ j_o}_{m_on_o}(\lambda) \prod^{I}_{i=1}  D^{*j_i}_{n'_im'_i}(\lambda)=\sum_x x^{*}_{m_1 \cdots m_O, n'_1\cdots n'_I}\; x_{n_1 \cdots n_O, m'_1 \cdots m'_I},
\label{int group}
\ee
where $x_{n_1 \cdots n_O, m'_1 \cdots m'_I}$ are the $SU(2)$ intertwiners between $I$ incoming and $O$ outgoing representations, respectively.
The projector \eqref{gauge}  restricts the $SU(2)$ functionals to be gauge invariant with coefficients
 \be
 <\Gamma,\{j_e\},\{x_v\}|\psi>= \psi_{j_e,x_v} =\prod_{v\in\Gamma} {x_{v}} \cdot   \prod_{e\in\Gamma} \psi^{j_{e}}_{mn}, 
 \ee
 where the gauge invariant basis elements are of the form
 \be
<h |\Gamma,\{j_e\},\{x_v\}>=\prod_{v\in\Gamma} {x_{v}}\cdot   \prod_{e\in\Gamma}D^{j_{e}}(h_{e})_{mn}. 
\label{spinnet solite} \ee
As we have seen in the previous section the imposition of the quantum constraint $\chi^2=0$ reduces the allowed SU(2) representations on the links to be of the kind $D^{j}_{jj}(h)$.
 Let's focus on a single vertex $v$  with $I$ ingoing links $e_i$ and $O$ outgoing links $e_o$, respectively, such that $s(e_o)=t(e_i)=v \quad \forall i,o$: if we apply the projector  \eqref{gauge} acting on $v$ to the quantum reduced basis elements (\ref{reduced basis elements}) we get:
 \be
 \begin{split}
 &P_{\mathcal{G}}  \prod^{O}_{o=1} {}^o\!D^{j_o}_{j_o j_o}(h_{e_o}) \prod^{I}_{i=1} {}^i\!D^{j_i}_{j_i j_i}(h_{e_i})= \int \; d\lambda_{v} \prod^{O}_{o=1} {}^{o}D^{j_o}_{j_o \alpha_o}(\lambda_{s(e_o)})\; {}^{o}\!D^{j_o}_{\alpha_o \beta_o}(h_{e_o})\;  rest_{\beta_oj_o}\ \; {rest'}_{j_o\beta'_i} \prod^{I}_{i=1} {}^{i}\!D^{j_i}_{ \beta'_i\alpha'_i}(h_{e_i}) {}^{i}D^{j_i}_{\alpha'_i j_i}(\lambda^{-1}_{t(e_i)})\; \\
 &=  \int \; d\lambda_v \prod^{O}_{o=1}  {D^{j}}^{-1}_{j_o\gamma_o}(\vec{u_o}) D^{j}_{\gamma_o \delta_o}(\lambda_v) {D^{j}}_{\delta_o \alpha_o}(\vec{u_o}) \;\;{}^{o}D^{j_o}_{\alpha_o \beta_o}(h_{e_o})  \;  {rest}_{\beta_oj_o} \\
& \quad \quad {rest'}_{j_i\beta'_i}\; \prod^{I}_{i=1}  \!{}^{i}D^{j_i}_{\beta'_i \alpha'_i}(h_{e_i})\;\; {D^{j}}^{-1}_{ \alpha'_i\delta'_i}(\vec{u_i})  D^{j}_{ \delta'_i\gamma'_i}(\lambda^{-1}_v) \ {D^{j}}_{\gamma'_i j_i}(\vec{u_i})  =\\
&= \sum_{x_v}{x^{*}_{v_s}}_{,  \gamma_1 \cdots \gamma_O, \gamma'_1 \cdots \gamma'_I }\;{x_{v_s}}_{, \delta_1... \delta_O, \delta'_1... \delta'_I} \,\prod^{O}_{o=1}  {D^{j}}^{-1}_{j_o\gamma_o}(\vec{u_o}) \;  {D^{j}}_{\delta_o \alpha_o}(\vec{u_o})\;\; {}^{o}\!D^{j_o}_{\alpha_o \beta_o}(h_{e_o})  {rest}_{\beta_oj_o} \\
&\prod^{I}_{i=1} \; {rest'}_{j_i\beta'_i}   \;  {D^{j}}^{-1}_{ \alpha'_i \delta'_i}(\vec{u_i})\; {D^{j}}_{\gamma'_i j_i}(\vec{u_i}) \;\;{}^{i}\!D^{j_i}_{ \beta'_i \alpha'_i}(h_{e_i}), 
 \end{split}
\label{proiettore completo}
 \ee
 where $rest$  ($rest'$) indicate the part of the holonomy whose final (initial) index transforms under gauge transformation with a group element $\tilde{\lambda}\neq\lambda_v$ and in the second and third equality we used the equations \eqref{D proiettate} and \eqref{int group} respectively.
 The previous expression can be reformulated introducing a Livine-Speziale coherent intertwiner \cite{ls} $|{\bf j_O}, \vec{{\bf u}}_O,{\bf j_I}, \vec{{\bf u}}_I  > \in \prod^{O}_{o} H^{j_o} \otimes \prod^{I}_{i} H^{*j_i}$ adapted to incoming and outgoing edges: 
 \be
|{\bf j_O}, \vec{{\bf u}}_O,{\bf j_I}, \vec{{\bf u}}_I  >= |{\bf j_O}, \vec{{\bf u}}_O>\otimes  <{\bf j_I}, \vec{{\bf u}}_I|=\int \;d\lambda \prod_{o=1}^{O} \lambda^{-1} | j_o, {\vec{u_o}}> \otimes  \prod_{i=1}^{I} < j_i, \vec{u_i}|{\lambda},
 \ee
 and noting that its projection on the usual Intertwiner base $|{\bf j_O},{\bf j_I},{\bf x} \rangle ={x^{*}_{v_s}}_{,  m_1 \cdots m_O, m'_1 \cdots m'_I }  \prod^{O}_{o} | j_o, m_o> \otimes  \prod^{I}_{i} < j_i, m'_i|$ with $|{\bf j_O},{\bf j_I},{\bf x} \rangle \in \prod^{O}_{o} H^{j_o} \otimes \prod^{I}_{i} H^{*j_i}$ is exactly the coefficient appearing in \eqref{proiettore completo}:
 \be
<{\bf j_O}, \vec{{\bf u}}_O,{\bf j_I}, \vec{{\bf u}}_I  | {\bf j_O},{\bf j_I}, {\bf x}_v >= {x^{*}_{v_s}}_{,  \gamma_1 \cdots \gamma_O, \gamma'_1 \cdots \gamma'_I }\prod^{O}_{o=1}  {D^{j}}^{-1}_{j_o\gamma_o}(\vec{u_o}) \prod^{I}_{i=1}  {D^{j}}_{\gamma'_i j_i}(\vec{u_i}), 
\label{reduced intertwiners}
 \ee
 or equivalently:
 \be
 \begin{split}
 &P_{\mathcal{G}}  \prod^{O}_{o=1} {}^o\!D^{j_o}_{j_o j_o}(h_{e_o}) \prod^{I}_{i=1} {}^i\!D^{j_i}_{j_i j_i}(h_{e_i})= \\
  &=\sum_{x_v}<{\bf j_O}, \vec{{\bf u}}_O,{\bf j_I}, \vec{{\bf u}}_I  | {\bf j_O},{\bf j_I}, {\bf x}_{v} >\;{x_{v_s}}_{, \delta_1... \delta_O, \delta'_1... \delta'_I} \,\prod^{O}_{o=1}   {D^{j}}_{\delta_o \alpha_o}(\vec{u_o}) \;\;{}^{o}\!D^{j_o}_{\alpha_o \beta_o}(h_{e_o}) \text{rest}_{\beta_oj_o}
\prod^{I}_{i=1}   \;  {D^{j}}^{-1}_{ \alpha'_i \delta'_i}(\vec{u_i})\;\; \text{rest'}_{j_i\beta'_i} {}^{i}\!D^{j_i}_{ \beta'_i \alpha'_i}(h_{e_i}), 
 \end{split}
\label{proiettore completo 2}
 \ee
thus we see that the gauge invariant projector brings us out of the space of reduced holonomies. This was expected since the Gauss constraint $\mathcal{G}$  that generates the $SU(2)$ transformations doesn't commute with the second class constraint $\chi=0$ imposed weakly. Our class of states can then be selected asking that the states averaged over $\mathcal{G}$ now also satisfy the constraint $\chi=0$; 
to ensure this condition it's enough to select the maximum weight spin in the sum over $\alpha_o$ and $\alpha'_i$ inside the expression \eqref{proiettore completo 2}
 \be
 \begin{split}
 &[ P_{\mathcal{G}}  \prod^{O}_{o=1} {}^o\!D^{j_o}_{j_o j_o}(h_{e_o}) \prod^{I}_{i=1} {}^i\!D^{j_i}_{j_i j_i}(h_{e_i})]_R= \\
& \sum_x  <{\bf j_O}, \vec{{\bf u}}_O,{\bf j_I}, \vec{{\bf u}}_I  | {\bf j_O},{\bf j_I}, {\bf x}_v >\;   
 <{ {\bf j_O},{\bf j_I}, {\bf x}_v|\bf j_O}, \vec{{\bf u}}_O,{\bf j_I}, \vec{{\bf u}}_I   >\;
  \prod^{O}_{o=1}   \;{}^{o}\!D^{j_o}_{j_o \beta_o}(h_{e_o}) \text{rest}_{\beta_oj_o}
\prod^{I}_{i=1}   \; \text{rest'}_{j_i\beta'_i} {}^{i}\!D^{j_i}_{ \beta'_i j_i}(h_{e_i}). 
 \end{split}
\label{proiettore completo 3}
 \ee

The previous equation can then be seen as the replacement of the usual projector on the gauge invariant states of the full theory $P_{\mathcal{G}}: \mathcal{H}^{kin} \rightarrow \;^{\mathcal{G}}\mathcal{H}^{kin}$  with it's reduced version $P_{\mathcal{G},\chi}:  \mathcal{H}^{kin} \rightarrow \; ^{\mathcal{G}} \mathcal{H}^{R}$ where $P_{\mathcal{G},\chi}=P^{\dag}_{\chi}\; P_{\mathcal{G}} P_{\chi}$ is given by the composition of the Gauss Projector with the projector on the quantum reduced space with $P_\chi: \mathcal{H}^{kin} \rightarrow \mathcal{H}^{R}$.

The reduced basis states will then be of the form:
\be
<h|\Gamma, j_e, x_v \rangle_R= \prod_{v\in\Gamma} <{\bf j_l}, {\bf x}_v|{\bf j_l},  \vec{{\bf u}}_l >  \cdot \prod_{e\in\Gamma} \;{}^l\!D^{j_{e_l}}_{j_{e_l} j_{e_l}}(h_{e_l}),
\label{base finale}
\ee
$l$ denoting ingoing and outgoing directions of the links $e_l$ with tangent vectors $u_l$ in $v$,  while $<{\bf j_l}, {\bf x}_v|{\bf j_l},  \vec{{\bf u}}_l >$ a short hand notation for the generic reduced intertwiner of the kind \eqref{reduced intertwiners}. The contraction now is just standard multiplication according to the orientation and connectivity of the holonomies.

The expansion of the projected spinnetwork \eqref{fine projected} on this base is then:

\be
{}_{R}<\Gamma, j_e, x_v | \psi> = \prod_{v\in\Gamma} < \vec{{\bf u}}_l, {\bf j_l}, |{\bf j_l}, {\bf x}_v >\cdot \;  \prod_{e\in\Gamma} \; {}^{l}\psi^{j_{e_l}}_{e_l}\;,
\ee
with ${}^{l}\psi^{j_{e_l}}_{e_l}=<j,\vec{u_l}|\psi^j|j,\vec{u_l}>$.
 
What's about the scalar product?
This is induced from the one of the full theory  {\it i.e.}
\be
<\Gamma, j_e, x_v |\Gamma', j'_e, x'_v >=\delta_{\Gamma,\Gamma'}\; \delta_{j_e,j'_e}\; \delta_{x_v,x_v'}.
\label{AL}
\ee
In fact looking at a single edge we see that (\ref{AL}) is based on the orthogonality relation

\be
\int d\lambda\; D^{*j_1}_{ab}(\lambda)\; D^{j_2}_{cd}(\lambda)=\frac{1}{d_{j_1}} \delta_{j_1,j_2} \delta_{ac}\; \delta_{bd},
\ee
that naturally induces on the reduced basis elements:
\be
\int d\lambda\; D^{*j_1}_{j_1j_1}(\lambda)\; D^{j_2}_{j_2j_2}(\lambda)=\frac{1}{d_{j_1}} \delta_{j_1,j_2},
\ee
equivalent up to a scaling to the $U(1)$ scalar product along each edge.
However the reduced states $|\Gamma', j'_e, x'_v >$ are not anymore orthogonal respect to the intertwiner because
\be
{}_R<\Gamma, j_e, x_v |\Gamma', j'_e, x'_v >_R=\delta_{\Gamma,\Gamma'}\; \delta_{j_e,j'_e}\; \prod_{v\in\Gamma} \prod_{e\in\Gamma}  <{\bf j_l}, \vec{{\bf u}}_l | {\bf j_l} , {\bf x_v}><{\bf j_l}, {\bf x'_v}|{\bf j_l}, \vec{{\bf u}}_l >,
\ee
 and we need to employ an orthonormalization procedure as the Gram-Schmidt one.
 
 It's interesting to note that using the resolution of the identity in terms of coherent states:
\be
\mathcal{I}_j=\sum_m |j,m><j,m|=d_j \int_{SU(2)} d\lambda\; |j,\lambda><j,\lambda|=d_j \int_{\mathcal{S}^2} d \vec{u} |j,\vec{u}><j,\vec{u}|\quad,
\ee
where $|j,\lambda>$ are coherent states defined as $|j,\lambda>=\lambda|j,j>$ and $|j,\vec{u}>$, with $\vec{u}$ unit vectors on the sphere $\mathcal{S}^2$, are proportional to $|j,\lambda>$ up to a phase that drops in the integral, one finds
\be
\delta_{ab}=<j,a |\mathcal{I}_j |j,b>=d_j \int_{\mathcal{S}^2} d \vec{u} <j,a | j,\vec{u}><j,\vec{u}|j,b>.
\ee 
Using the previous expression then a generic basis element $D^j_{mn}(g)$ of  $\mathcal{H}^{kin}$ can be written as:
\be
\begin{split}
&D^j_{mn}=\sum_{ab} \delta_{ma} D^j_{ab} \delta_{bn}=\\
&=\sum_{ab} d^2_j \int_{\mathcal{S}^2} d \vec{u} <j,m | j,\vec{u}><j,\vec{u}|j,a> D^j_{ab} \int_{\mathcal{S}^2} d \vec{u'} <j,b | j,\vec{u'}><j,\vec{u'}|j,n>.
\label{Dcoerente}
\end{split}
\ee
This expression is now useful to infer the form of the reduced states basis of ${}^{\mathcal{G}} \mathcal{H}^{R}$.

The quantum constraint $\chi^2$ in fact will act at the end point (the conjugate condition will hold at the starting point) of the holonomy as:
\be
\begin{split}
&\hat{\chi}^2  D^j(g) |j,\vec{u}> =D^j(g) (\tau^2 - (\vec{e_l} \cdot\vec{\tau})^2 ) |j,\vec{u}>= D^j(g) (j (j+1) - (\vec{e_l} \cdot\vec{\tau})^2 ) |j,\vec{u}>
\end{split}
\ee
and using the property of the coherent states $\vec{v} \cdot\vec{\tau}|j,\vec{v}>=j |j,\vec{v}>$ we see that if and only if \be
\vec{e_l}=\vec{u}
\label {condizione}
\ee
the basis elements will satisfy $\hat{\chi}^2  D^j(g) |j,\vec{u}>=0$ in the appropriate limit. Then looking at equation \eqref{Dcoerente} we see that the index $m,n$ will be connected to the usual intertwiners at the starting and endpoints of the holonomies for gauge invariant states in $^{\mathcal{G}} \mathcal{H} $ of the kind \eqref{spinnet solite}, but all the terms in the two integrals for which the condition \eqref{condizione} does not hold, are not solutions of the constraint $\hat{\chi}^2=0$. The reduced holonomies to be connected to the standard intertwiners are then :
\be
\tilde{g}={}^{R} D^j_{mn}(g)= d^2_j <j,m | j,\vec{e_l}><j,\vec{e_l}| D^j(g)| j,\vec{e_l}><j,\vec{e_l}|j,n>.
\label{reduced with free indexes}
\ee

The previous expression gives a simple rule to build states in ${}^{\mathcal{G}} \mathcal{H}^{R}$: it is enough to connect expressions \eqref{reduced with free indexes} instead of the usual $D^j_{mn}(g)$ to the standard intertwiners.

These states, which are defined over  cuboidal lattices with 6-valent intertwiners, are suitable to describe a quantum Universe in the case of the inhomogeneous extension of the Bianchi I model.

\subsection{Intertwiners}

In the previous section we determined the reduction implied by the condition (\ref{chi}). This procedure is well-grounded because only interior edge points have been considered, while holonomies transform under gauge transformations at boundary points only. Hence, the underlying SU(2) gauge structure becomes manifest at vertexes. We are going to evaluate the expression of intertwiners adapted to the reduced holonomies $\tilde{h}_{e_l}$ which are the quantum version of \eqref{rhol} \emph{i.e} ${}^{R}\hat{h}=\tilde{h}$ . 

The basic scheme of recoupling theory is given by 3-valent intertwiners. Let us consider a 3-valent vertex with with two edges $e_1$ and $e_2$ incoming and $e_3$ outcoming, with associated $SU(2)$ irreps $j_1$, $j_2$ and $j_3$, respectively. The SU(2) intertwiner is given by the Clebsch-Gordan coefficients or equivalently by 3j symbols (equipped with the $1j$ ``metric tensor'') \cite{BrinkSatchler68} , such that the full vertex reads 
\begin{equation}
D^{j_1}_{n'n}(h_{e_1})D^{j_2}_{p'p}(h_{e_2})C^{j_3 m}_{j_1 j_2 n p}D^{j_3}_{mm'}(h_{e_3} ).
\end{equation}
where the repeated magnetic indexes are all summed and taken in a fixed basis, for example the one that diagonalizes $\tau^3$.

In order to find out U(1) irreps out of SU(2) ones, holonomies must be written in the basis that stabilize the direction $e_l$ and the restriction to the representations with maximum magnetic numbers must be considered. In particular is convenient to introduce a graphical representation for the expression \eqref{reduced with free indexes} 

\be
\begin{split}
&{}^{R} D^j(g)_{mn}= d^2_j <j,m | j,\vec{e_l}><j,\vec{e_l}| D^j(g)| j,\vec{e_l}><j,\vec{e_l}|j,n>=\\
&=\begin{array} {c} 
\ifx\JPicScale\undefined\def\JPicScale{1.1}\fi
\psset{unit=\JPicScale mm}
\psset{linewidth=0.3,dotsep=1,hatchwidth=0.3,hatchsep=1.5,shadowsize=1,dimen=middle}
\psset{dotsize=0.7 2.5,dotscale=1 1,fillcolor=black}
\psset{arrowsize=1 2,arrowlength=1,arrowinset=0.25,tbarsize=0.7 5,bracketlength=0.15,rbracketlength=0.15}
\begin{pspicture}(0,0)(68,8)
\psline{|*-}(52,3)(48,3)
\rput(35,8){$\scr{j}$}
\rput{0}(35,3){\psellipse[](0,0)(3,-3)}
\rput(35,3){$g$}
\psline[fillstyle=solid]{-|}(12,3)(16,3)
\pspolygon[](6,6)(12,6)(12,0)(6,0)
\psline[fillstyle=solid](2,3)(6,3)
\rput(9,3){$\scr{R_{\vec{e_l}}}$}
\psline[fillstyle=solid](64,3)(68,3)
\pspolygon[](58,6)(64,6)(64,0)(58,0)
\psline[fillstyle=solid]{|*-}(54,3)(58,3)
\rput(61,3){$\scr{R^{-1}_{\vec{e_l}}}$}
\pspolygon[](42,6)(48,6)(48,0)(42,0)
\rput(45,3){$\scr{R_{\vec{e_l}}}$}
\psline[fillstyle=solid](38,3)(42,3)
\psline{}(32,3)(28,3)
\pspolygon[](22,6)(28,6)(28,0)(22,0)
\rput(25,3){$\scr{R^{-1}_{\vec{e_l}}}$}
\psline[fillstyle=solid]{|-}(18,3)(22,3)
\rput(9,8){$\scr{j}$}
\rput(61,8){$\scr{j}$}
\end{pspicture}
\end{array}
\end{split}
\label{Dridottamn}
\ee

where the solid lines represent the identity in the base $|j,m>$ that diagonalizes $\tau^3$,

\be
\delta_{ab}=\begin{array} {c} 
\ifx\JPicScale\undefined\def\JPicScale{1.1}\fi
\psset{unit=\JPicScale mm}
\psset{linewidth=0.3,dotsep=1,hatchwidth=0.3,hatchsep=1.5,shadowsize=1,dimen=middle}
\psset{dotsize=0.7 2.5,dotscale=1 1,fillcolor=black}
\psset{arrowsize=1 2,arrowlength=1,arrowinset=0.25,tbarsize=0.7 5,bracketlength=0.15,rbracketlength=0.15}
\begin{pspicture}(0,0)(10,8)
\psline[fillstyle=solid](2,3)(7,3)
\rput(4,2){$\scr{j}$}
\end{pspicture}
\end{array},
\ee
the projection on the maximum magnetic number is given by 
\be
<j,j|=\begin{array} {c} 
\ifx\JPicScale\undefined\def\JPicScale{1.1}\fi
\psset{unit=\JPicScale mm}
\psset{linewidth=0.3,dotsep=1,hatchwidth=0.3,hatchsep=1.5,shadowsize=1,dimen=middle}
\psset{dotsize=0.7 2.5,dotscale=1 1,fillcolor=black}
\psset{arrowsize=1 2,arrowlength=1,arrowinset=0.25,tbarsize=0.7 5,bracketlength=0.15,rbracketlength=0.15}
\begin{pspicture}(0,0)(10,8)
\psline[fillstyle=solid]{|-}(2,3)(7,3)
\rput(4,2){$\scr{j}$}
\end{pspicture}
\end{array}
\quad
\text{and} 
\quad
|j,j>=
\begin {array} {c}
\ifx\JPicScale\undefined\def\JPicScale{1.1}\fi
\psset{unit=\JPicScale mm}
\psset{linewidth=0.3,dotsep=1,hatchwidth=0.3,hatchsep=1.5,shadowsize=1,dimen=middle}
\psset{dotsize=0.7 2.5,dotscale=1 1,fillcolor=black}
\psset{arrowsize=1 2,arrowlength=1,arrowinset=0.25,tbarsize=0.7 5,bracketlength=0.15,rbracketlength=0.15}
\begin{pspicture}(0,0)(10,8)
\psline[fillstyle=solid]{-|}(2,3)(7,3)
\rput(4,2){$\scr{j}$}
\end{pspicture}
\end{array}
\ee

and the group elements are represented by boxes or circles depending if they represent the Wigner matrix  of specific fixed rotation $R(e_l)$ that move the $e_z$ axis to the $e_l$ axis, selecting the desired $U(1)_l$ subgroup, or a generic $SU(2)$ Wigner matrix $D(g)$ in the $|j,m>$ base, respectively:

\be
\begin{split}
D(R(\vec{e_l}))^j_{mn}=\begin{array} {c} 
\ifx\JPicScale\undefined\def\JPicScale{1.1}\fi
\psset{unit=\JPicScale mm}
\psset{linewidth=0.3,dotsep=1,hatchwidth=0.3,hatchsep=1.5,shadowsize=1,dimen=middle}
\psset{dotsize=0.7 2.5,dotscale=1 1,fillcolor=black}
\psset{arrowsize=1 2,arrowlength=1,arrowinset=0.25,tbarsize=0.7 5,bracketlength=0.15,rbracketlength=0.15}
\begin{pspicture}(0,0)(28,8)
\rput(9,8){$\scr{j}$}
\psline[fillstyle=solid](12,3)(16,3)
\pspolygon[](6,6)(12,6)(12,0)(6,0)
\psline[fillstyle=solid](2,3)(6,3)
\rput(9,3){$\scr{R_{\vec{e_l}}}$}
\end{pspicture}
\end{array}
\end{split}
\ee

\be
D^j_{mn}(g)=\begin{array} {c} 
\ifx\JPicScale\undefined\def\JPicScale{1.1}\fi
\psset{unit=\JPicScale mm}
\psset{linewidth=0.3,dotsep=1,hatchwidth=0.3,hatchsep=1.5,shadowsize=1,dimen=middle}
\psset{dotsize=0.7 2.5,dotscale=1 1,fillcolor=black}
\psset{arrowsize=1 2,arrowlength=1,arrowinset=0.25,tbarsize=0.7 5,bracketlength=0.15,rbracketlength=0.15}
\begin{pspicture}(0,0)(50,8)
\rput(7,8){$\scr{j}$}
\rput{0}(7,3){\psellipse[](0,0)(3,-3)}
\rput(7,3){$g$}
\psline[fillstyle=solid](10,3)(14,3)
\psline{}(0,3)(4,3)
\end{pspicture}
\end{array}.
\ee

With this notation the 3-valent intertwiner from which any higher valence one (our theory for the Bianchi I model prescribes nodes at most 6-valent) can be represented as

\be
 <j_1,j_2, j_3 , x_v |  j_1,j_2,j_3, e_1, e_2, e_3 >=\begin{array} {c}
\ifx\JPicScale\undefined\def\JPicScale{1}\fi
\psset{unit=\JPicScale mm}
\psset{linewidth=0.3,dotsep=1,hatchwidth=0.3,hatchsep=1.5,shadowsize=1,dimen=middle}
\psset{dotsize=0.7 2.5,dotscale=1 1,fillcolor=black}
\psset{arrowsize=1 2,arrowlength=1,arrowinset=0.25,tbarsize=0.7 5,bracketlength=0.15,rbracketlength=0.15}
\begin{pspicture}(0,0)(40,25)
\psline{|-}(0,0)(8,4)
\psline{-|*}(20,10)(20,25)
\psline{-|*}(32,4)(40,0)
\pspolygon[](8,8)(12,8)(12,4)(8,4)
\pspolygon[](28,8)(32,8)(32,4)(28,4)
\psline(12,6)(20,10)
\psline(20,10)(28,6)
\rput(10,6){$\scr{R_{e_1}}$}
\rput(30,6){$\scr{R_{e_2}}$}
\rput(22,18){$j_3$}
\rput(17,6){$j_1$}
\rput(24,6){$j_2$}
\end{pspicture}
\end{array}
\label{3-valent reduced}
\ee

where the 3-valent node is the usual 3-j symbol contracted with the SU(2) coherent states in the three directions $e_1,e_2,e_3$ . The explicit value of the function \eqref{3-valent reduced}  can then  be computed using the values of the Wigner matrix for a rotation parametrized  for example  by the Euler angles $(\alpha,\beta,\gamma)$ that brings the vector $(0,0,1)$ to the vector $e_l$.
The Wigner matrices are then given by
\be
D^j_{m,m'}(\alpha,\beta,\gamma)=e^{im\alpha} d^j_{mm'}(\beta)e^{im'\gamma},
\ee 
where $d^j_{mm'}(\beta)$ is the Wigner function given in Appendix \ref{app}.
In particular for the cubical lattice we are interested to the vectors $e_3=e_z=(0,0,1)$, $e_2=e_y=(0,1,0)$ and $e_1=e_x=(1,0,0)$ and the rotation matrices appearing in \eqref{3-valent reduced}  are 
given by $D^j_{m,m'}(-\frac{\pi}{2},\frac{\pi}{2},\frac{\pi}{2}):=R_y$ and $D^j_{m,m'}(0,\frac{\pi}{2},0):=R_x$. In fact the two matrices rotate the $z$ axis respectively into the $y$ and the $x$ direction.

This graphical machinery can now be used to introduce a reduced recoupling theory (see Appendix \ref{app2}) out of the $SU(2)$ one and to compute the action of the scalar constraint.

\subsection{Geometric operators}
In reduced Hilbert space ${}^\mathcal{G}\mathcal{H}^R$, the following relation defining the action of fluxes on basis elements holds

\be
{}_R<e_l, j_{e_l}| \hat{E}_i(S^k)|e_l, j_{e_l}>_{R}=<e_l, j_{e_l}|P_{\chi} \hat{E}_i(S^k) P_{\chi}|e_l, j_{e_l}>= -i \; 8\pi\gamma l_P^2  o(e_i,S^i)\; \delta_{ik} \delta_{kl}\; j_{e_l},
\ee

where $e_l$ indicates the edge $e$ in direction $l$.

Henceforth, geometric operators are developed starting from the action of reduced fluxes $P_{\chi} \hat{E}_i(S^k) P_{\chi}={}^{R}\!\hat{E}_i(S^k)$ in reduced Hilbert space. Therefore, the area operator operator along a surface $S^i$ is given by
\be
{}^{R}\!\hat{A}[S^i]=\int \sqrt{{}^{R}\!\hat{E}_i(S^i){}^{R}\!\hat{E}_i(S^i)}dudv,
\ee

$u,v$ being a proper parametrization of the surface $S^i$. The expression above can be regularized as in the full theory \cite{discr}, and at the end its action is non-vanishing only on  
$|e_i,j_{e_i}>$ for $e_i\cap S\neq\emptyset$ giving
\be
{}^{R}\!\hat{A}[S^i]|e_i,j_{e_i}>=8\pi l_P^2\gamma j_{e_i}|e_i,j_{e_i}>.
\ee

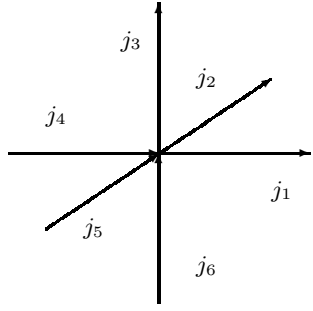
\begin{figure}
$
\begin{array} {c}
\ifx\JPicScale\undefined\def\JPicScale{1}\fi
\unitlength \JPicScale mm
\begin{picture}(100,70)(-30,20)
\linethickness{0.3mm}
\put(30,50){\line(1,0){20}}
\put(50,50){\vector(1,0){0.12}}
\linethickness{0.3mm}
\put(30,50){\line(0,1){20}}
\put(30,70){\vector(0,1){0.12}}
\linethickness{0.3mm}
\multiput(30,50)(0.18,0.12){83}{\line(1,0){0.18}}
\put(45,60){\vector(3,2){0.12}}
\linethickness{0.3mm}
\put(10,50){\line(1,0){20}}
\put(30,50){\vector(1,0){0.12}}
\linethickness{0.3mm}
\put(30,30){\line(0,1){20}}
\put(30,50){\vector(0,1){0.12}}
\linethickness{0.3mm}
\multiput(15,40)(0.18,0.12){83}{\line(1,0){0.18}}
\put(30,50){\vector(3,2){0.12}}

\put(45,45){\makebox(0,0)[cl]{$j_1$}}

\put(35,60){\makebox(0,0)[cl]{$j_2$}}

\put(25,65){\makebox(0,0)[cl]{$j_3$}}

\put(15,55){\makebox(0,0)[cl]{$j_4$}}

\put(20,40){\makebox(0,0)[cl]{$j_5$}}

\put(35,35){\makebox(0,0)[cl]{$j_6$}}
\end{picture}
\end{array}
$
\nonumber
\caption{6-valent vertex.} \label{6v}
\end{figure}

In the same way, the action of the volume operator can be defined from reduced fluxes and its expression gets an enormous simplification with respect to the full theory \cite{vol1,vol2} thanks to the reduction of $SU(2)$ group elements to $U(1)_i$ ones. Let us consider the volume of a region $\Omega$ containing only one vertex $v$ (the extension to regions containing more than one vertex is straightforward), the operator $V(\Omega)$ reads
\be
{}^{R}\!\hat{V}(\Omega)=\sum_{a,b,c,i,k,l}\int d^3x\sqrt{|\frac{1}{3!}\epsilon_{abc}\epsilon^{ikl}{}^{R}\!\hat{E}_i^a {}^{R}\!\hat{E}_k^b{}^{R}\!\hat{E}_l^c|},
\ee

and the action on a trivalent node $|e_1,e_2,e_3,j_{e_1},j_{e_2},j_{e_3},x_v \rangle$ can be regularized by introducing reduced fluxes over $S^i$ $i=1,2,3$ with $S^1\cap S^2\cap S^3=v$ as follows
\begin{eqnarray}
{}^{R}\!\hat{V}(\Omega)|e_1,e_2,e_3,j_{e_1},j_{e_2},j_{e_3},x_v \rangle=\nonumber\\=\int d^3x\sqrt{|\sum_{i,k,l,m,n,p}\frac{1}{3!}\epsilon_{mnp}\epsilon^{ikl}{}^{R}\! \hat{E}_i(S^m) {}^{R}\!\hat{E}_k(S^n){}^{R}\! \hat{E}_l(S^p)|}|e_1,e_2,e_3,j_{e_1},j_{e_2},j_{e_3},x_v \rangle\nonumber\\=
\int d^3x\sqrt{|\sum_{i,k,l}\frac{1}{3!}(\epsilon_{ikl})^2{}^{R}\! \hat{E}_i(S^i) {}^{R}\! \hat{E}_k(S^k){}^{R}\!\hat{E}_l(S^l)|}|e_1,e_2,e_3,j_{e_1},j_{e_2},j_{e_3},x_v \rangle=
\nonumber\\=\int d^3x\sqrt{|{}^{R}\!\hat{E}_1(S^1) {}^{R}\!\hat{E}_2(S^2){}^{R}\!\hat{E}_3(S^3)|}|e_1,e_2,e_3,j_{e_1},j_{e_2},j_{e_3},x_v \rangle=\nonumber\\
=\sqrt{|j_{e_1}j_{e_2}j_{e_3}o(e_1,S^1)o(e_2,S^2)o(e_3,S^3)|}|e_1,e_2,e_3,j_{e_1},j_{e_2},j_{e_3},x_v \rangle,
\end{eqnarray}

where in the second and third lines we used the fact that ${}^{R}\!\hat{E}_i(S^m)$ is nonvanishing only if $i=m$ and the commutativity of ${}^{R}\!\hat{E}_i(S^i)$ and ${}^{R}\!\hat{E}_l(S^l)$. Henceforth, \emph{the action of the volume operator is diagonal in the basis \eqref{base finale}}.
In the case of a generic vertex, the expression above should be summed over all the $e_1$, $e_2$ and $e_3$ emanating from $v$, 
and it does not depend explicitly on the intertwiner structure. For the 6-valent vertex in figure \ref{6v} by choosing the orientation of $S^1$, $S^2$ and $S^3$ such that their normals are parallel to $e_1$, $e_2$ and $e_3$, respectively, the volume becomes 
\begin{eqnarray}
{}^{R}\!\hat{V}(\Omega)|e_1,e_2,e_3,e_4,e_5,e_6, j_1,j_2,j_3,j_4,j_5,j_6, x_v\rangle=\nonumber\\=(8\pi\gamma l_P^2)^{3/2}\sqrt{(j_1+j_4)(j_2+j_5)(j_3+j_6)}
|e_1,e_2,e_3,e_4,e_5,e_6, j_1,j_2,j_3,j_4,j_5,j_6, x_v\rangle.
\end{eqnarray}

\section{Hamiltonian Constraint}\label{Hmatrix}
The superHamiltonian operator can be consistently regularized in the reduced Hilbert space, starting from the expression of the full theory. In this respect, let us restrict our attention to the Euclidean constraint (\ref{H_E_N0}) and let us adapt Thiemann regularization procedure \cite{qsd} to the states of the reduced theory by considering only cubic cells. 
Hence, let us develop a cubulation $C$ of the manifold $\Sigma$ adapted to the Graph $\Gamma$ underlying the cubical lattices over which our reduced cylindrical functions are defined.
For each pair of links $e_i$ and $e_j$ incident at a node $\vgraph$ of $\Gamma$ we choose a semi-analytic arcs $a_{ij}$ that respect the lattice structure, {\it i.e.} such that the end points $s_{e_i},s_{e_j}$ are interior points of $e_i,e_j$, respectively, and  $a_{ij}\cap\Gamma=\{s_{e_i},s_{e_j}\}$. The arc $s_i$ is the segment of $e_i$($e_j$) from $\vgraph$ to $s_i$($s_j$), while $s_{i}$, $s_{j}$ and $a_{ij}$ generate a rectangle $\alpha_{ij} := s_{i} \circ a_{ij} \circ s_j^{-1}$.
  Three (non-planar) links define a cube and we get a complete cubulation $C$ of the spatial manifold by summing over all the incident edges at a given node and over all nodes. Now we can decompose \eqref{H_E_N0} obtaining the expressions \eqref{Ham_T10} and \eqref{Hm_delta:classical10} adapted to $C$ simply with the replacement  $\Delta \rightarrow \ba
 \begin{tikzpicture}
\pgfmathsetmacro{\cubex}{0.15}
\pgfmathsetmacro{\cubey}{0.15}
\pgfmathsetmacro{\cubez}{0.15}
\draw (0,0,0) -- ++(-\cubex,0,0) -- ++(0,-\cubey,0) -- ++(\cubex,0,0) -- cycle;
\draw (0,0,0) -- ++(0,0,-\cubez) -- ++(0,-\cubey,0) -- ++(0,0,\cubez) -- cycle;
\draw (0,0,0) -- ++(-\cubex,0,0) -- ++(0,0,-\cubez) -- ++(\cubex,0,0) -- cycle;
\end{tikzpicture}
\ea$
.
 
We are now interested in  implementing the action of the operator (\ref{Hm_delta:quantum}) via an operator ${}^{R}\hat{H}$ defined on $^{\mathcal{G}} \mathcal{H}^{R}$: a convenient way of constructing it is to replace in the expression (\ref{Hm_delta:quantum}) quantum holonomies and fluxes with the ones acting on the reduced space as follows

\be
{}^{R}\hat{H}^m_{\cube}[N]:= N(\vgraph) C(m) \,  \, \epsilon^{ijk} \,
   \mathrm{Tr}\Big[{}^{R}\hat{h}^{(m)}_{\alpha_{ij}} {}^{R}\hat{h}^{(m)-1}_{s_{k}} \big[{}^R\hat{h}^{(m)}_{s_{k}},{}^{R}\!\hat{V}\big]\Big]. 
   \label{Hridotto}
\ee
The lattice spacing of the cubulation $C$ that acts as a regularization parameter and it can be changed via a reduced diffeomorphisms. Hence, on reduced s-knot states there exists a suitable operator topology in which the regulator can be safely removed as in full LQG. Therefore, \emph{the action of the superHamiltonian operator can be regularized in the reduced Hilbert space}. 
We proceed in the next section to the explicit computation of the matrix elements of  \eqref{Hridotto} on three valent nodes.

\subsection {Reduced Hamiltonian on three valent node}

The reduced Hamiltonian ${}^{R}\hat{H}^m_{\cube}[N]$ acts on reduced states as the operator (\ref{Hm_delta:quantum}) does on ordinary spinnetwork states.
The difference is that now the holonomies are of the kind \eqref{reduced basis elements} or equivalently of the kind \eqref{Dridottamn} considering the node structure and consequently one has to recouple them using the rules contained in Appendix \ref{app2} .
Here we study ${}^{R}\hat{H}^m_{\cube}[N]$ acting on a three-valent node; in the following we neglect the value of the lapse function and the constants analyzing the operator

\be
{}^{R}\hat{H}^m_{\cube}=\epsilon^{ijk} \,
   \mathrm{Tr}\Big[{}^{R}\hat{h}^{(m)}_{\alpha_{ij}} {}^{R}\hat{h}^{(m)-1}_{s_{k}} {}^{R}\!\hat{V} {}^R\hat{h}^{(m)}_{s_{k}}\Big],
   \label{Hcalcolo}
\ee
because, as is the full theory, due to the presence of $\epsilon^{ijk}$ this is the only non vanishing term in the commutator.
Choosing a three valent vertex state $|v_3\rangle_R=|e_x,e_y,e_z, j_x, j_y, j_z , x_v \rangle_R$ with outgoing edges $e_x, e_y, e_z$, the Hamiltonian ${}^{R}\hat{H}^m_{\cube} |v_3 \rangle_R$ is the sum of three terms  ${}^{R}\hat{H}^m_{\cube} |v_3 \rangle_R=\sum_{k=1}^3{}^{R}\hat{H}^m_{k\cube} |v_3 \rangle_R$ where $k=1,2,3$ for $s_k\in e_x,e_y. e_z$ respectively. $\hat{H}^m_{3\cube} |v_3 \rangle_R$ acts first by multiplication with the holonomy on the right of \eqref{Hcalcolo} producing:

\be
\hat{h}^{(m)}_{s_{z}} |v_3\rangle_R\; = \hat{h}^{(m)}_{s_{z}}
\ba
\ifx\JPicScale\undefined\def\JPicScale{0.7}\fi
\ifx\JPicScale\undefined\def\JPicScale{1}\fi
\psset{unit=\JPicScale mm}
\psset{linewidth=0.3,dotsep=1,hatchwidth=0.3,hatchsep=1.5,shadowsize=1,dimen=middle}
\psset{dotsize=0.7 2.5,dotscale=1 1,fillcolor=black}
\psset{arrowsize=1 2,arrowlength=1,arrowinset=0.25,tbarsize=0.7 5,bracketlength=0.15,rbracketlength=0.15}
\begin{pspicture}(0,65)(90,140)
\psline{|*-}(19,78)(21.83,80.83)
\rput(17,81){$\scr{j_x}$}
\rput{44.4}(28.5,87.5){\psellipse[](0,0)(2.21,-2.12)}
\rput(24,91){$\scr{h_{x}}$}
\psline[fillstyle=solid]{-|}(42.83,101.83)(40,99)
\pspolygon[](46,103)(44,101)(42,103)(44,105)
\psline[fillstyle=solid](51,110)(44.84,103.84)
\rput(40,104){$\scr{R_x}$}
\psline[fillstyle=solid]{|*-}(16.83,75.83)(14,73)
\rput(10,74){$\scr{R^{-1}_x}$}
\rput(21,85){$\scr{R_x}$}
\psline[fillstyle=solid](26.83,85.83)(24,83)
\psline(30,89)(32.82,91.82)
\pspolygon[](36,93)(34,91)(32,93)(34,95)
\rput(32,97){$\scr{R^{-1}_x}$}
\psline[fillstyle=solid]{|-}(38,97)(35.17,94.17)
\rput(43,108){$\scr{j_x}$}
\rput(13,77){$\scr{j_x}$}
\psline{|*-}(51,137)(51,133)
\rput{0}(51,130){\psellipse[](0,0)(3,-3)}
\rput(51,130){$\scr{h_{z}}$}
\psline[fillstyle=solid]{-|}(51,110)(51,120)
\psline[fillstyle=solid]{|*-}(51,139)(51,143)
\psline{-|}(51,127)(51,123)
\rput(46,141){$\scr{j_z}$}
\pspolygon[](25,82)(23,80)(21,82)(23,84)
\psline[fillstyle=solid](12,71)(9.17,68.17)
\pspolygon[](15,72)(13,70)(11,72)(13,74)
\psline{|*-}(83,78)(80.17,80.83)
\rput(84,82){$\scr{j_y}$}
\rput(76,91){$\scr{h_{y}}$}
\psline[fillstyle=solid]{-|}(59.17,101.83)(62,99)
\pspolygon[](58,105)(60,103)(58,101)(56,103)
\rput(62,104){$\scr{R_y}$}
\psline[fillstyle=solid]{|*-}(85.17,75.83)(88,73)
\rput(95,73){$\scr{R^{-1}_y}$}
\rput(79,86){$\scr{R_y}$}
\psline[fillstyle=solid](75.17,85.83)(78,83)
\psline(72,89)(69.18,91.82)
\pspolygon[](68,95)(70,93)(68,91)(66,93)
\rput(72,96){$\scr{R^{-1}_y}$}
\psline[fillstyle=solid]{|-}(64,97)(66.83,94.17)
\rput(57,107){$\scr{j_y}$}
\rput(90,77){$\scr{j_y}$}
\pspolygon[](79,84)(81,82)(79,80)(77,82)
\psline[fillstyle=solid](90,71)(92.83,68.17)
\pspolygon[](89,74)(91,72)(89,70)(87,72)
\rput{44.4}(73.44,87.44){\psellipse[](0,0)(2.21,-2.12)}
\psline[fillstyle=solid](51,110)(57.16,103.84)
\rput(48,115){$\scr{j_z}$}
\end{pspicture}
\ea
=
\ee
\be
=d_{j_z+m}
\ba
\ifx\JPicScale\undefined\def\JPicScale{0.7}\fi
\psset{unit=\JPicScale mm}
\psset{linewidth=0.3,dotsep=1,hatchwidth=0.3,hatchsep=1.5,shadowsize=1,dimen=middle}
\psset{dotsize=0.7 2.5,dotscale=1 1,fillcolor=black}
\psset{arrowsize=1 2,arrowlength=1,arrowinset=0.25,tbarsize=0.7 5,bracketlength=0.15,rbracketlength=0.15}
\begin{pspicture}(0,70)(90,138)
\psline{|*-}(19,78)(21.83,80.83)
\rput(17,81){$\scr{j_x}$}
\rput{44.4}(28.5,87.5){\psellipse[](0,0)(2.21,-2.12)}
\rput(24,91){$\scr{h_{x}}$}
\psline[fillstyle=solid]{-|}(42.83,101.83)(40,99)
\pspolygon[](46,103)(44,101)(42,103)(44,105)
\psline[fillstyle=solid](51,110)(44.84,103.84)
\rput(40,104){$\scr{R_x}$}
\psline[fillstyle=solid]{|*-}(16.83,75.83)(14,73)
\rput(10,74){$\scr{R^{-1}_x}$}
\rput(21,85){$\scr{R_x}$}
\psline[fillstyle=solid](26.83,85.83)(24,83)
\psline(30,89)(32.82,91.82)
\pspolygon[](36,93)(34,91)(32,93)(34,95)
\rput(32,97){$\scr{R^{-1}_x}$}
\psline[fillstyle=solid]{|-}(38,97)(35.17,94.17)
\rput(43,108){$\scr{j_x}$}
\rput(13,77){$\scr{j_x}$}
\psline{|*-}(51,137)(51,133)
\rput{0}(51,130){\psellipse[](0,0)(3,-3)}
\rput(51,130){$\scr{h_{z}}$}
\psline[fillstyle=solid]{-|}(51,110)(51,120)
\psline[fillstyle=solid]{|*-}(51,139)(51,145)
\psline{-|}(51,127)(51,123)
\rput(46,144){$\scr{j_z}$}
\pspolygon[](25,82)(23,80)(21,82)(23,84)
\psline[fillstyle=solid](12,71)(9.17,68.17)
\pspolygon[](15,72)(13,70)(11,72)(13,74)
\psline{|*-}(83,78)(80.17,80.83)
\rput(84,82){$\scr{j_y}$}
\rput(76,91){$\scr{h_{y}}$}
\psline[fillstyle=solid]{-|}(59.17,101.83)(62,99)
\pspolygon[](58,105)(60,103)(58,101)(56,103)
\rput(62,104){$\scr{R_y}$}
\psline[fillstyle=solid]{|*-}(85.17,75.83)(88,73)
\rput(95,73){$\scr{R^{-1}_y}$}
\rput(79,86){$\scr{R_y}$}
\psline[fillstyle=solid](75.17,85.83)(78,83)
\psline(72,89)(69.18,91.82)
\pspolygon[](68,95)(70,93)(68,91)(66,93)
\rput(72,96){$\scr{R^{-1}_y}$}
\psline[fillstyle=solid]{|-}(64,97)(66.83,94.17)
\rput(57,107){$\scr{j_y}$}
\rput(90,77){$\scr{j_y}$}
\pspolygon[](79,84)(81,82)(79,80)(77,82)
\psline[fillstyle=solid](90,71)(92.83,68.17)
\pspolygon[](89,74)(91,72)(89,70)(87,72)
\rput{44.4}(73.44,87.44){\psellipse[](0,0)(2.21,-2.12)}
\psline[fillstyle=solid](51,110)(57.16,103.84)
\rput(46,112){$\scr{j_z}$}
\psline(51,115)(55,115)
\psline(51,142)(55,142)
\rput(46,118){$\scr{j_z+m}$}
\rput(57,117){$\scr{m}$}
\rput(58,141){$\scr{m}$}
\rput(45,139){$\scr{j_z+m}$}
\end{pspicture}.
\ea
\ee
Then the Volume acts diagonally multipling by $\sqrt{j_x\;j_y\;(j_z+m)}$ and the last operator ${}^{R}\hat{h}^{(m)}_{\alpha_{ij}} {}^{R}\hat{h}^{(m)-1}_{s_{k}}$ attaches the inverse holonomy and the loop $\alpha_{xy}$. We get
\be
\mathrm{Tr}\Big[{}^{R}\hat{h}^{(m)}_{\alpha_{xy}} {}^{R}\hat{h}^{(m)-1}_{s_{z}} {}^{R}\!\hat{V} {}^R\hat{h}^{(m)}_{s_{z}}\Big] |v_3\rangle_R
=
\sqrt{j_x\;j_y\;(j_z+1)}  \frac{(d_{j_z})^2}{d_{j_z+m}}
\ba
\ifx\JPicScale\undefined\def\JPicScale{0.7}\fi
\ifx\JPicScale\undefined\def\JPicScale{1}\fi
\psset{unit=\JPicScale mm}
\psset{linewidth=0.3,dotsep=1,hatchwidth=0.3,hatchsep=1.5,shadowsize=1,dimen=middle}
\psset{dotsize=0.7 2.5,dotscale=1 1,fillcolor=black}
\psset{arrowsize=1 2,arrowlength=1,arrowinset=0.25,tbarsize=0.7 5,bracketlength=0.15,rbracketlength=0.15}
\begin{pspicture}(0,0)(90,138)
\psline{|*-}(19,78)(21.83,80.83)
\rput(17,81){$\scr{j_x}$}
\rput{45}(28.5,87.5){\psellipse[](0,0)(2.21,-2.12)}
\rput(24,91){$\scr{h_{x}}$}
\psline[fillstyle=solid]{-|}(42.83,101.83)(40,99)
\pspolygon[](46,103)(44,101)(42,103)(44,105)
\psline[fillstyle=solid](51,110)(44.84,103.84)
\rput(40,104){$\scr{R_x}$}
\psline[fillstyle=solid]{|*-}(16.83,75.83)(14,73)
\rput(10,74){$\scr{R^{-1}_x}$}
\rput(21,85){$\scr{R_x}$}
\psline[fillstyle=solid](26.83,85.83)(24,83)
\psline(30,89)(32.82,91.82)
\pspolygon[](36,93)(34,91)(32,93)(34,95)
\rput(32,97){$\scr{R^{-1}_x}$}
\psline[fillstyle=solid]{|-}(38,97)(35.17,94.17)
\rput(43,108){$\scr{j_x}$}
\rput(13,77){$\scr{j_x}$}
\psline{|*-}(51,137)(51,133)
\rput(44,129){$\scr{j_z}$}
\rput{0}(51,130){\psellipse[](0,0)(3,-3)}
\rput(51,130){$\scr{h_{z}}$}
\psline[fillstyle=solid]{-|}(51,110)(51,120)
\psline[fillstyle=solid]{|*-}(51,139)(51,143)
\psline{-|}(51,127)(51,123)
\rput(45,115){$\scr{j_z}+m$}
\rput(46,141){$\scr{j_z}$}
\psline(51,114)(53,114)
\psline(51,117)(54,117)
\pspolygon[](25,82)(23,80)(21,82)(23,84)
\psline[fillstyle=solid](12,71)(6,65)
\pspolygon[](15,72)(13,70)(11,72)(13,74)
\psline{|*-}(83,78)(80.17,80.83)
\rput(84,82){$\scr{j_y}$}
\rput(76,91){$\scr{h_{y}}$}
\psline[fillstyle=solid]{-|}(59.17,101.83)(62,99)
\pspolygon[](58,105)(60,103)(58,101)(56,103)
\rput(62,104){$\scr{R_y}$}
\psline[fillstyle=solid]{|*-}(85.17,75.83)(88,73)
\rput(95,73){$\scr{R^{-1}_y}$}
\rput(79,86){$\scr{R_y}$}
\psline[fillstyle=solid](75.17,85.83)(78,83)
\psline(72,89)(69.18,91.82)
\pspolygon[](68,95)(70,93)(68,91)(66,93)
\rput(72,96){$\scr{R^{-1}_y}$}
\psline[fillstyle=solid]{|-}(64,97)(66.83,94.17)
\rput(57,107){$\scr{j_y}$}
\rput(90,77){$\scr{j_y}$}
\pspolygon[](79,84)(81,82)(79,80)(77,82)
\psline[fillstyle=solid](90,71)(96,65)
\pspolygon[](89,74)(91,72)(89,70)(87,72)
\rput{43.83}(73.44,87.44){\psellipse[](0,0)(2.21,-2.12)}
\psline{|*-}(42,93)(39.17,90.17)
\rput(26,74){$\scr{m}$}
\rput{45}(32.5,83.5){\psellipse[](0,0)(2.21,-2.12)}
\rput(37,84){$\scr{h_{x}}$}
\psline[fillstyle=solid]{-|}(18.17,69.17)(21,72)
\pspolygon[](15,68)(17,70)(19,68)(17,66)
\psline[fillstyle=solid](14,65)(16.16,67.16)
\rput(21,67){$\scr{R^{-1}_x}$}
\psline[fillstyle=solid]{|*-}(44.17,95.17)(47,98)
\rput(48,94){$\scr{R_x}$}
\rput(42,88){$\scr{R^{-1}_x}$}
\psline[fillstyle=solid](34.17,85.17)(37,88)
\psline(31,82)(28.18,79.18)
\pspolygon[](25,78)(27,80)(29,78)(27,76)
\rput(31,78){$\scr{R_x}$}
\psline[fillstyle=solid]{|-}(23,74)(25.83,76.83)
\rput(14,56){$\scr{m}$}
\rput(51,104){$\scr{m}$}
\pspolygon[](36,89)(38,91)(40,89)(38,87)
\psline[fillstyle=solid](49,100)(53,104)
\pspolygon[](46,99)(48,101)(50,99)(48,97)
\psline{|*-}(60,93)(62.83,90.17)
\rput(74,74){$\scr{m}$}
\rput(63,82){$\scr{h^{-1}_{y}}$}
\psline[fillstyle=solid]{-|}(83.83,69.17)(81,72)
\pspolygon[](85,66)(83,68)(85,70)(87,68)
\psline[fillstyle=solid](89,64)(85.84,67.16)
\rput(81,67){$\scr{R_y}$}
\psline[fillstyle=solid]{|*-}(57.83,95.17)(55,98)
\rput(55,94){$\scr{R_y^{-1}}$}
\rput(60,88){$\scr{R_y}$}
\psline[fillstyle=solid](67.83,85.17)(65,88)
\psline(71,82)(73.82,79.18)
\pspolygon[](75,76)(73,78)(75,80)(77,78)
\rput(70,77){$\scr{R^{-1}_y}$}
\psline[fillstyle=solid]{|-}(79,74)(76.17,76.83)
\rput(80,62){$\scr{m}$}
\rput(56,115){$\scr{m}$}
\pspolygon[](64,87)(62,89)(64,91)(66,89)
\pspolygon[](54,97)(52,99)(54,101)(56,99)
\rput{45}(69.56,83.56){\psellipse[](0,0)(2.21,-2.12)}
\psline[fillstyle=solid](53,104)(53,114)
\psline[fillstyle=solid](54,101)(54,117)
\psline[border=0.3,fillstyle=solid](51,110)(57.16,103.84)
\rput(49,112){$\scr{j_z}$}
\rput(47,120){$\scr{j_z}$}
\psline{|*-}(42,37)(39.17,39.83)
\rput(43,41){$\scr{j_y}$}
\rput(35,50){$\scr{h_{y}}$}
\psline[fillstyle=solid]{-|}(18.17,60.83)(21,58)
\pspolygon[](17,64)(19,62)(17,60)(15,62)
\rput(21,63){$\scr{R_y}$}
\psline[fillstyle=solid]{|*-}(44.17,34.83)(47,32)
\rput(44,28){$\scr{R^{-1}_y}$}
\rput(38,45){$\scr{R_y}$}
\psline[fillstyle=solid](34.17,44.83)(37,42)
\psline(31,48)(28.18,50.82)
\pspolygon[](27,54)(29,52)(27,50)(25,52)
\rput(31,55){$\scr{R^{-1}_y}$}
\psline[fillstyle=solid]{|-}(23,56)(25.83,53.17)
\rput(49,36){$\scr{j_y}$}
\pspolygon[](38,43)(40,41)(38,39)(36,41)
\psline[fillstyle=solid](49,30)(51.83,27.17)
\pspolygon[](48,33)(50,31)(48,29)(46,31)
\rput{43.83}(32.44,46.44){\psellipse[](0,0)(2.21,-2.12)}
\psline[fillstyle=solid](14,65)(16.16,62.84)
\psline{|*-}(80,55)(77.17,52.17)
\rput(64,36){$\scr{m}$}
\rput{45}(70.5,45.5){\psellipse[](0,0)(2.21,-2.12)}
\rput(75,44){$\scr{h^{-1}_{x}}$}
\psline[fillstyle=solid]{-|}(56.17,31.17)(59,34)
\pspolygon[](53,30)(55,32)(57,30)(55,28)
\psline[fillstyle=solid](52,27)(54.16,29.16)
\rput(58,26){$\scr{R_x}$}
\psline[fillstyle=solid]{|*-}(82.17,57.17)(85,60)
\rput(88,56){$\scr{R^{-1}_x}$}
\rput(80,50){$\scr{R_x}$}
\psline[fillstyle=solid](72.17,47.17)(75,50)
\psline(69,44)(66.18,41.18)
\pspolygon[](63,40)(65,42)(67,40)(65,38)
\rput(71,38){$\scr{R^{-1}_x}$}
\psline[fillstyle=solid]{|-}(61,36)(63.83,38.83)
\pspolygon[](74,51)(76,53)(78,51)(76,49)
\psline[fillstyle=solid](87,62)(89,64)
\pspolygon[](84,61)(86,63)(88,61)(86,59)
\end{pspicture}
\ea
\ee
Reduced recoupling (see Appendix \ref{app2}) on the links gives:
\be
\mathrm{Tr}\Big[{}^{R}\hat{h}^{(m)}_{\alpha_{xy}} {}^{R}\hat{h}^{(m)-1}_{s_{z}} {}^{R}\!\hat{V} {}^R\hat{h}^{(m)}_{s_{z}}\Big] |v_3\rangle_R
=
\sqrt{j_x\;j_y\;(j_z+1)}  \frac{(d_{j_z})^2}{d_{j_z+m}} d_{j_x+m} d_{j_y-m}
\ba
\ifx\JPicScale\undefined\def\JPicScale{0.7}\fi
\psset{unit=\JPicScale mm}
\psset{linewidth=0.3,dotsep=1,hatchwidth=0.3,hatchsep=1.5,shadowsize=1,dimen=middle}
\psset{dotsize=0.7 2.5,dotscale=1 1,fillcolor=black}
\psset{arrowsize=1 2,arrowlength=1,arrowinset=0.25,tbarsize=0.7 5,bracketlength=0.15,rbracketlength=0.15}
\begin{pspicture}(0,0)(97,143)
\psline{|*-}(19,78)(21.83,80.83)
\rput(13,81){$\scr{j_x+m}$}
\rput{45}(28.5,87.5){\psellipse[](0,0)(2.21,-2.12)}
\rput(24,91){$\scr{h_{x}}$}
\psline[fillstyle=solid]{-|}(42.83,101.83)(40,99)
\pspolygon[](46,103)(44,101)(42,103)(44,105)
\psline[fillstyle=solid](51,110)(44.84,103.84)
\rput(40,104){$\scr{R_x}$}
\psline[fillstyle=solid]{|*-}(16.83,75.83)(14,73)
\rput(10,74){$\scr{R^{-1}_x}$}
\rput(21,85){$\scr{R_x}$}
\psline[fillstyle=solid](26.83,85.83)(24,83)
\psline(30,89)(32.82,91.82)
\pspolygon[](36,93)(34,91)(32,93)(34,95)
\rput(32,97){$\scr{R^{-1}_x}$}
\psline[fillstyle=solid]{|-}(38,97)(35.17,94.17)
\rput(43,108){$\scr{j_x}$}
\rput(11,77){$\scr{j_x+m}$}
\psline{|*-}(51,137)(51,133)
\rput(44,129){$\scr{j_z}$}
\rput{0}(51,130){\psellipse[](0,0)(3,-3)}
\rput(51,130){$\scr{h_{z}}$}
\psline[fillstyle=solid]{-|}(51,110)(51,120)
\psline[fillstyle=solid]{|*-}(51,139)(51,143)
\psline{-|}(51,127)(51,123)
\rput(45,115){$\scr{j_z}+m$}
\rput(46,141){$\scr{j_z}$}
\psline(51,114)(53,114)
\psline(51,117)(54,117)
\pspolygon[](25,82)(23,80)(21,82)(23,84)
\psline[fillstyle=solid](12,71)(6,65)
\pspolygon[](15,72)(13,70)(11,72)(13,74)
\psline{|*-}(83,78)(80.17,80.83)
\rput(86,82){$\scr{j_y-m}$}
\rput(76,91){$\scr{h_{y}}$}
\psline[fillstyle=solid]{-|}(59.17,101.83)(62,99)
\pspolygon[](58,105)(60,103)(58,101)(56,103)
\rput(62,104){$\scr{R_y}$}
\psline[fillstyle=solid]{|*-}(85.17,75.83)(88,73)
\rput(95,73){$\scr{R^{-1}_y}$}
\rput(79,86){$\scr{R_y}$}
\psline[fillstyle=solid](75.17,85.83)(78,83)
\psline(72,89)(69.18,91.82)
\pspolygon[](68,95)(70,93)(68,91)(66,93)
\rput(72,96){$\scr{R^{-1}_y}$}
\psline[fillstyle=solid]{|-}(64,97)(66.83,94.17)
\rput(57,107){$\scr{j_y}$}
\rput(90,77){$\scr{j_y-m}$}
\pspolygon[](79,84)(81,82)(79,80)(77,82)
\psline[fillstyle=solid](90,71)(96,65)
\pspolygon[](89,74)(91,72)(89,70)(87,72)
\rput{43.83}(73.44,87.44){\psellipse[](0,0)(2.21,-2.12)}
\psline[fillstyle=solid](41,100)(47,98)
\rput(46,94){$\scr{R_x}$}
\rput(14,60){$\scr{m}$}
\rput(51,104){$\scr{m}$}
\psline[fillstyle=solid](49,100)(53,104)
\pspolygon[](46,99)(48,101)(50,99)(48,97)
\psline[fillstyle=solid](61,100)(55,98)
\rput(80,64){$\scr{m}$}
\rput(56,115){$\scr{m}$}
\pspolygon[](54,97)(52,99)(54,101)(56,99)
\psline[fillstyle=solid](53,104)(53,114)
\psline[fillstyle=solid](54,101)(54,117)
\psline[border=0.3,fillstyle=solid](51,110)(57.16,103.84)
\rput(49,112){$\scr{j_z}$}
\rput(47,120){$\scr{j_z}$}
\psline{|*-}(42,39)(39.17,41.83)
\rput(43,43){$\scr{m}$}
\rput(35,52){$\scr{h_{y}}$}
\psline[fillstyle=solid]{-|}(18.17,62.83)(21,60)
\pspolygon[](17,66)(19,64)(17,62)(15,64)
\rput(21,65){$\scr{R_y}$}
\psline[fillstyle=solid]{|*-}(44.17,36.83)(47,34)
\rput(44,30){$\scr{R^{-1}_y}$}
\rput(38,47){$\scr{R_y}$}
\psline[fillstyle=solid](34.17,46.83)(37,44)
\psline(31,50)(28.18,52.82)
\pspolygon[](27,56)(29,54)(27,52)(25,54)
\rput(31,57){$\scr{R^{-1}_y}$}
\psline[fillstyle=solid]{|-}(23,58)(25.83,55.17)
\rput(49,38){$\scr{m}$}
\pspolygon[](38,45)(40,43)(38,41)(36,43)
\psline[fillstyle=solid](49,32)(51.83,29.17)
\pspolygon[](48,35)(50,33)(48,31)(46,33)
\rput{43.83}(32.44,48.44){\psellipse[](0,0)(2.21,-2.12)}
\psline[fillstyle=solid](11,70)(16.16,64.84)
\psline{|*-}(80,57)(77.17,54.17)
\rput(64,38){$\scr{m}$}
\rput{45}(70.5,47.5){\psellipse[](0,0)(2.21,-2.12)}
\rput(75,46){$\scr{h^{-1}_{x}}$}
\psline[fillstyle=solid]{-|}(56.17,33.17)(59,36)
\pspolygon[](53,32)(55,34)(57,32)(55,30)
\psline[fillstyle=solid](52,29)(54.16,31.16)
\rput(58,28){$\scr{R_x}$}
\psline[fillstyle=solid]{|*-}(82.17,59.17)(85,62)
\rput(88,58){$\scr{R^{-1}_x}$}
\rput(80,52){$\scr{R_x}$}
\psline[fillstyle=solid](72.17,49.17)(75,52)
\psline(69,46)(66.18,43.18)
\pspolygon[](63,42)(65,44)(67,42)(65,40)
\rput(71,40){$\scr{R^{-1}_x}$}
\psline[fillstyle=solid]{|-}(61,38)(63.83,40.83)
\pspolygon[](74,53)(76,55)(78,53)(76,51)
\psline[fillstyle=solid](87,64)(92,69)
\pspolygon[](84,63)(86,65)(88,63)(86,61)
\rput(35,100){$\scr{j_x+m}$}
\rput(67,101){$\scr{j_y-m}$}
\rput(97,67){$\scr{j_y}$}
\rput(5,68){$\scr{j_x}$}
\rput(56,94){$\scr{R^{-1}_y}$}
\end{pspicture}
\ea
\ee
The previous expression can then be simplified moving the box using the invariance of the intertwiners at the central node and using  $SU(2)$ recoupling theory we get (see %
\cite{mio1,io e antonia}):
\be
\begin{split}
&\mathrm{Tr}\Big[{}^{R}\hat{h}^{(m)}_{\alpha_{xy}} {}^{R}\hat{h}^{(m)-1}_{s_{z}} {}^{R}\!\hat{V} {}^R\hat{h}^{(m)}_{s_{z}}\Big] |v_3\rangle_R
=\\
=&\sqrt{j_x\;j_y\;(j_z+1)}  \frac{(d_{j_z})^2}{d_{j_z+m}} d_{j_x+m} d_{j_y-m} (-1)^{3m} \tinysixj{j_{x+m}} {j_y} {j_{z+m}} {j_z} {m} {j_x} \tinysixj{j_{x+m}} {j_{y-m}} {j_{z}} {m} {j_{z+m}} {j_y}
\ba
\ifx\JPicScale\undefined\def\JPicScale{0.7}\fi
\psset{unit=\JPicScale mm}
\psset{linewidth=0.3,dotsep=1,hatchwidth=0.3,hatchsep=1.5,shadowsize=1,dimen=middle}
\psset{dotsize=0.7 2.5,dotscale=1 1,fillcolor=black}
\psset{arrowsize=1 2,arrowlength=1,arrowinset=0.25,tbarsize=0.7 5,bracketlength=0.15,rbracketlength=0.15}
\begin{pspicture}(0,0)(97,143)
\psline{|*-}(19,78)(21.83,80.83)
\rput(13,81){$\scr{j_x+m}$}
\rput{45}(28.5,87.5){\psellipse[](0,0)(2.21,-2.12)}
\rput(24,91){$\scr{h_{x}}$}
\psline[fillstyle=solid]{-|}(42.83,101.83)(40,99)
\pspolygon[](46,103)(44,101)(42,103)(44,105)
\psline[fillstyle=solid](51,110)(44.84,103.84)
\rput(40,104){$\scr{R_x}$}
\psline[fillstyle=solid]{|*-}(16.83,75.83)(14,73)
\rput(10,74){$\scr{R^{-1}_x}$}
\rput(21,85){$\scr{R_x}$}
\psline[fillstyle=solid](26.83,85.83)(24,83)
\psline(30,89)(32.82,91.82)
\pspolygon[](36,93)(34,91)(32,93)(34,95)
\rput(32,97){$\scr{R^{-1}_x}$}
\psline[fillstyle=solid]{|-}(38,97)(35.17,94.17)
\rput(11,77){$\scr{j_x+m}$}
\psline{|*-}(51,137)(51,133)
\rput(44,129){$\scr{j_z}$}
\rput{0}(51,130){\psellipse[](0,0)(3,-3)}
\rput(51,130){$\scr{h_{z}}$}
\psline[fillstyle=solid]{-|}(51,110)(51,120)
\psline[fillstyle=solid]{|*-}(51,139)(51,143)
\psline{-|}(51,127)(51,123)
\rput(46,141){$\scr{j_z}$}
\pspolygon[](25,82)(23,80)(21,82)(23,84)
\psline[fillstyle=solid](12,71)(6,65)
\pspolygon[](15,72)(13,70)(11,72)(13,74)
\psline{|*-}(83,78)(80.17,80.83)
\rput(84,82){$\scr{j_y-m}$}
\rput(76,91){$\scr{h_{y}}$}
\psline[fillstyle=solid]{-|}(59.17,101.83)(62,99)
\pspolygon[](58,105)(60,103)(58,101)(56,103)
\rput(62,104){$\scr{R_y}$}
\psline[fillstyle=solid]{|*-}(85.17,75.83)(88,73)
\rput(95,73){$\scr{R^{-1}_y}$}
\rput(79,86){$\scr{R_y}$}
\psline[fillstyle=solid](75.17,85.83)(78,83)
\psline(72,89)(69.18,91.82)
\pspolygon[](68,95)(70,93)(68,91)(66,93)
\rput(72,96){$\scr{R^{-1}_y}$}
\psline[fillstyle=solid]{|-}(64,97)(66.83,94.17)
\rput(90,77){$\scr{j_y-m}$}
\pspolygon[](79,84)(81,82)(79,80)(77,82)
\psline[fillstyle=solid](90,71)(96,65)
\pspolygon[](89,74)(91,72)(89,70)(87,72)
\rput{43.83}(73.44,87.44){\psellipse[](0,0)(2.21,-2.12)}
\rput(14,60){$\scr{m}$}
\rput(80,64){$\scr{m}$}
\psline[border=0.3,fillstyle=solid](51,110)(57.16,103.84)
\rput(47,120){$\scr{j_z}$}
\psline{|*-}(42,39)(39.17,41.83)
\rput(43,43){$\scr{m}$}
\rput(35,52){$\scr{h_{y}}$}
\psline[fillstyle=solid]{-|}(18.17,62.83)(21,60)
\pspolygon[](17,66)(19,64)(17,62)(15,64)
\rput(21,65){$\scr{R_y}$}
\psline[fillstyle=solid]{|*-}(44.17,36.83)(47,34)
\rput(44,30){$\scr{R^{-1}_y}$}
\rput(38,47){$\scr{R_y}$}
\psline[fillstyle=solid](34.17,46.83)(37,44)
\psline(31,50)(28.18,52.82)
\pspolygon[](27,56)(29,54)(27,52)(25,54)
\rput(31,57){$\scr{R^{-1}_y}$}
\psline[fillstyle=solid]{|-}(23,58)(25.83,55.17)
\rput(49,38){$\scr{m}$}
\pspolygon[](38,45)(40,43)(38,41)(36,43)
\psline[fillstyle=solid](49,32)(51.83,29.17)
\pspolygon[](48,35)(50,33)(48,31)(46,33)
\rput{43.83}(32.44,48.44){\psellipse[](0,0)(2.21,-2.12)}
\psline[fillstyle=solid](11,70)(16.16,64.84)
\psline{|*-}(80,57)(77.17,54.17)
\rput(64,38){$\scr{m}$}
\rput{45}(70.5,47.5){\psellipse[](0,0)(2.21,-2.12)}
\rput(75,46){$\scr{h^{-1}_{x}}$}
\psline[fillstyle=solid]{-|}(56.17,33.17)(59,36)
\pspolygon[](53,32)(55,34)(57,32)(55,30)
\psline[fillstyle=solid](52,29)(54.16,31.16)
\rput(58,28){$\scr{R_x}$}
\psline[fillstyle=solid]{|*-}(82.17,59.17)(85,62)
\rput(88,58){$\scr{R^{-1}_x}$}
\rput(80,52){$\scr{R_x}$}
\psline[fillstyle=solid](72.17,49.17)(75,52)
\psline(69,46)(66.18,43.18)
\pspolygon[](63,42)(65,44)(67,42)(65,40)
\rput(71,40){$\scr{R^{-1}_x}$}
\psline[fillstyle=solid]{|-}(61,38)(63.83,40.83)
\pspolygon[](74,53)(76,55)(78,53)(76,51)
\psline[fillstyle=solid](87,64)(92,69)
\pspolygon[](84,63)(86,65)(88,63)(86,61)
\rput(41,108){$\scr{j_x+m}$}
\rput(58,108){$\scr{j_y-m}$}
\rput(97,67){$\scr{j_y}$}
\rput(5,68){$\scr{j_x}$}
\end{pspicture}
\ea
\end{split}
\ee

A similar calculations for the reversed loop $\alpha_{yx}$ leads to the final result:

\be
\begin{split}
&
\hat{H}^m_{3\cube} |v_3 \rangle_R |v_3\rangle_R= \sqrt{j_x\;j_y\;(j_z+1)}  \frac{(d_{j_z})^2}{d_{j_z+m}} d_{j_x+m} d_{j_y-m} (-1)^{3m} \tinysixj{j_{x+m}} {j_y} {j_{z+m}} {j_z} {m} {j_x} \tinysixj{j_{x+m}} {j_{y-m}} {j_{z}} {m} {j_{z+m}} {j_y}
\\
&\ba
\ifx\JPicScale\undefined\def\JPicScale{0.7}\fi
\psset{unit=\JPicScale mm}
\psset{linewidth=0.3,dotsep=1,hatchwidth=0.3,hatchsep=1.5,shadowsize=1,dimen=middle}
\psset{dotsize=0.7 2.5,dotscale=1 1,fillcolor=black}
\psset{arrowsize=1 2,arrowlength=1,arrowinset=0.25,tbarsize=0.7 5,bracketlength=0.15,rbracketlength=0.15}
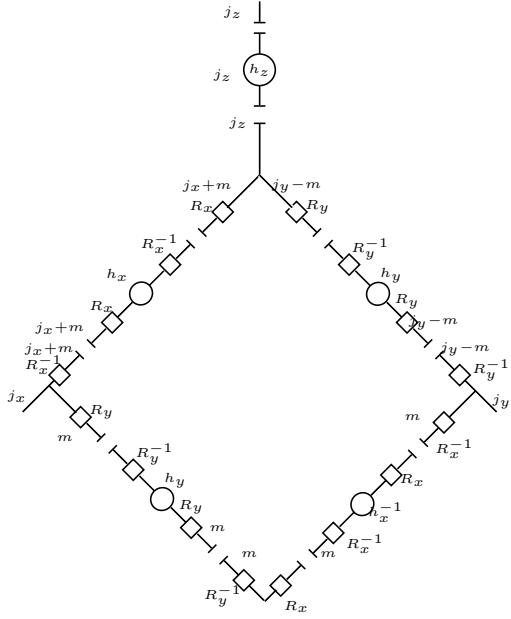
\begin{pspicture}(0,0)(97,143)
\psline{|*-}(19,78)(21.83,80.83)
\rput(13,81){$\scr{j_x+m}$}
\rput{45}(28.5,87.5){\psellipse[](0,0)(2.21,-2.12)}
\rput(24,91){$\scr{h_{x}}$}
\psline[fillstyle=solid]{-|}(42.83,101.83)(40,99)
\pspolygon[](46,103)(44,101)(42,103)(44,105)
\psline[fillstyle=solid](51,110)(44.84,103.84)
\rput(40,104){$\scr{R_x}$}
\psline[fillstyle=solid]{|*-}(16.83,75.83)(14,73)
\rput(10,74){$\scr{R^{-1}_x}$}
\rput(21,85){$\scr{R_x}$}
\psline[fillstyle=solid](26.83,85.83)(24,83)
\psline(30,89)(32.82,91.82)
\pspolygon[](36,93)(34,91)(32,93)(34,95)
\rput(32,97){$\scr{R^{-1}_x}$}
\psline[fillstyle=solid]{|-}(38,97)(35.17,94.17)
\rput(11,77){$\scr{j_x+m}$}
\psline{|*-}(51,137)(51,133)
\rput(44,129){$\scr{j_z}$}
\rput{0}(51,130){\psellipse[](0,0)(3,-3)}
\rput(51,130){$\scr{h_{z}}$}
\psline[fillstyle=solid]{-|}(51,110)(51,120)
\psline[fillstyle=solid]{|*-}(51,139)(51,143)
\psline{-|}(51,127)(51,123)
\rput(46,141){$\scr{j_z}$}
\pspolygon[](25,82)(23,80)(21,82)(23,84)
\psline[fillstyle=solid](12,71)(6,65)
\pspolygon[](15,72)(13,70)(11,72)(13,74)
\psline{|*-}(83,78)(80.17,80.83)
\rput(84,82){$\scr{j_y-m}$}
\rput(76,91){$\scr{h_{y}}$}
\psline[fillstyle=solid]{-|}(59.17,101.83)(62,99)
\pspolygon[](58,105)(60,103)(58,101)(56,103)
\rput(62,104){$\scr{R_y}$}
\psline[fillstyle=solid]{|*-}(85.17,75.83)(88,73)
\rput(95,73){$\scr{R^{-1}_y}$}
\rput(79,86){$\scr{R_y}$}
\psline[fillstyle=solid](75.17,85.83)(78,83)
\psline(72,89)(69.18,91.82)
\pspolygon[](68,95)(70,93)(68,91)(66,93)
\rput(72,96){$\scr{R^{-1}_y}$}
\psline[fillstyle=solid]{|-}(64,97)(66.83,94.17)
\rput(90,77){$\scr{j_y-m}$}
\pspolygon[](79,84)(81,82)(79,80)(77,82)
\psline[fillstyle=solid](90,71)(96,65)
\pspolygon[](89,74)(91,72)(89,70)(87,72)
\rput{43.83}(73.44,87.44){\psellipse[](0,0)(2.21,-2.12)}
\rput(14,60){$\scr{m}$}
\rput(80,64){$\scr{m}$}
\psline[border=0.3,fillstyle=solid](51,110)(57.16,103.84)
\rput(47,120){$\scr{j_z}$}
\psline{|*-}(42,39)(39.17,41.83)
\rput(43,43){$\scr{m}$}
\rput(35,52){$\scr{h_{y}}$}
\psline[fillstyle=solid]{-|}(18.17,62.83)(21,60)
\pspolygon[](17,66)(19,64)(17,62)(15,64)
\rput(21,65){$\scr{R_y}$}
\psline[fillstyle=solid]{|*-}(44.17,36.83)(47,34)
\rput(44,30){$\scr{R^{-1}_y}$}
\rput(38,47){$\scr{R_y}$}
\psline[fillstyle=solid](34.17,46.83)(37,44)
\psline(31,50)(28.18,52.82)
\pspolygon[](27,56)(29,54)(27,52)(25,54)
\rput(31,57){$\scr{R^{-1}_y}$}
\psline[fillstyle=solid]{|-}(23,58)(25.83,55.17)
\rput(49,38){$\scr{m}$}
\pspolygon[](38,45)(40,43)(38,41)(36,43)
\psline[fillstyle=solid](49,32)(51.83,29.17)
\pspolygon[](48,35)(50,33)(48,31)(46,33)
\rput{43.83}(32.44,48.44){\psellipse[](0,0)(2.21,-2.12)}
\psline[fillstyle=solid](11,70)(16.16,64.84)
\psline{|*-}(80,57)(77.17,54.17)
\rput(64,38){$\scr{m}$}
\rput{45}(70.5,47.5){\psellipse[](0,0)(2.21,-2.12)}
\rput(75,46){$\scr{h^{-1}_{x}}$}
\psline[fillstyle=solid]{-|}(56.17,33.17)(59,36)
\pspolygon[](53,32)(55,34)(57,32)(55,30)
\psline[fillstyle=solid](52,29)(54.16,31.16)
\rput(58,28){$\scr{R_x}$}
\psline[fillstyle=solid]{|*-}(82.17,59.17)(85,62)
\rput(88,58){$\scr{R^{-1}_x}$}
\rput(80,52){$\scr{R_x}$}
\psline[fillstyle=solid](72.17,49.17)(75,52)
\psline(69,46)(66.18,43.18)
\pspolygon[](63,42)(65,44)(67,42)(65,40)
\rput(71,40){$\scr{R^{-1}_x}$}
\psline[fillstyle=solid]{|-}(61,38)(63.83,40.83)
\pspolygon[](74,53)(76,55)(78,53)(76,51)
\psline[fillstyle=solid](87,64)(92,69)
\pspolygon[](84,63)(86,65)(88,63)(86,61)
\rput(41,108){$\scr{j_x+m}$}
\rput(58,108){$\scr{j_y-m}$}
\rput(97,67){$\scr{j_y}$}
\rput(5,68){$\scr{j_x}$}
\end{pspicture}
\ea
\\
&
- \sqrt{j_x\;j_y\;(j_z+1)}  \frac{(d_{j_z})^2}{d_{j_z+m}} d_{j_x-m} d_{j_y+m} (-1)^{3m} \tinysixj{j_{x}} {j_y+m} {j_{z+m}} {m} {j_z} {j_y} \tinysixj{j_{x}-m} {j_{y}+m} {j_{z}} {j_z+m} {m} {j_x}
\\
&\ba
\ifx\JPicScale\undefined\def\JPicScale{0.7}\fi
\psset{unit=\JPicScale mm}
\psset{linewidth=0.3,dotsep=1,hatchwidth=0.3,hatchsep=1.5,shadowsize=1,dimen=middle}
\psset{dotsize=0.7 2.5,dotscale=1 1,fillcolor=black}
\psset{arrowsize=1 2,arrowlength=1,arrowinset=0.25,tbarsize=0.7 5,bracketlength=0.15,rbracketlength=0.15}
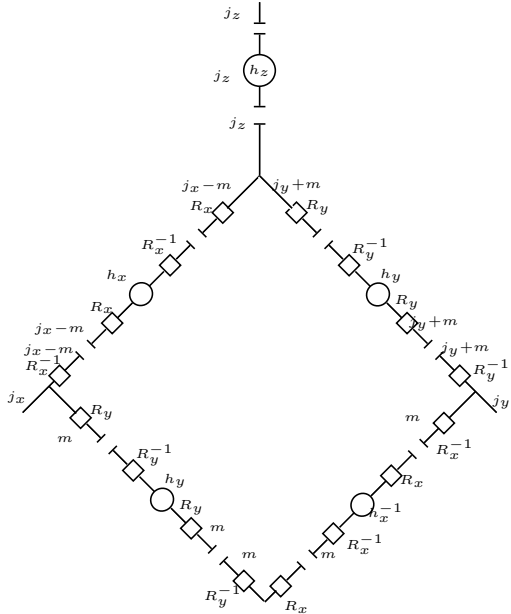
\begin{pspicture}(0,0)(97,143)
\psline{|*-}(19,78)(21.83,80.83)
\rput(13,81){$\scr{j_x-m}$}
\rput{45}(28.5,87.5){\psellipse[](0,0)(2.21,-2.12)}
\rput(24,91){$\scr{h_{x}}$}
\psline[fillstyle=solid]{-|}(42.83,101.83)(40,99)
\pspolygon[](46,103)(44,101)(42,103)(44,105)
\psline[fillstyle=solid](51,110)(44.84,103.84)
\rput(40,104){$\scr{R_x}$}
\psline[fillstyle=solid]{|*-}(16.83,75.83)(14,73)
\rput(10,74){$\scr{R^{-1}_x}$}
\rput(21,85){$\scr{R_x}$}
\psline[fillstyle=solid](26.83,85.83)(24,83)
\psline(30,89)(32.82,91.82)
\pspolygon[](36,93)(34,91)(32,93)(34,95)
\rput(32,97){$\scr{R^{-1}_x}$}
\psline[fillstyle=solid]{|-}(38,97)(35.17,94.17)
\rput(11,77){$\scr{j_x-m}$}
\psline{|*-}(51,137)(51,133)
\rput(44,129){$\scr{j_z}$}
\rput{0}(51,130){\psellipse[](0,0)(3,-3)}
\rput(51,130){$\scr{h_{z}}$}
\psline[fillstyle=solid]{-|}(51,110)(51,120)
\psline[fillstyle=solid]{|*-}(51,139)(51,143)
\psline{-|}(51,127)(51,123)
\rput(46,141){$\scr{j_z}$}
\pspolygon[](25,82)(23,80)(21,82)(23,84)
\psline[fillstyle=solid](12,71)(6,65)
\pspolygon[](15,72)(13,70)(11,72)(13,74)
\psline{|*-}(83,78)(80.17,80.83)
\rput(84,82){$\scr{j_y+m}$}
\rput(76,91){$\scr{h_{y}}$}
\psline[fillstyle=solid]{-|}(59.17,101.83)(62,99)
\pspolygon[](58,105)(60,103)(58,101)(56,103)
\rput(62,104){$\scr{R_y}$}
\psline[fillstyle=solid]{|*-}(85.17,75.83)(88,73)
\rput(95,73){$\scr{R^{-1}_y}$}
\rput(79,86){$\scr{R_y}$}
\psline[fillstyle=solid](75.17,85.83)(78,83)
\psline(72,89)(69.18,91.82)
\pspolygon[](68,95)(70,93)(68,91)(66,93)
\rput(72,96){$\scr{R^{-1}_y}$}
\psline[fillstyle=solid]{|-}(64,97)(66.83,94.17)
\rput(90,77){$\scr{j_y+m}$}
\pspolygon[](79,84)(81,82)(79,80)(77,82)
\psline[fillstyle=solid](90,71)(96,65)
\pspolygon[](89,74)(91,72)(89,70)(87,72)
\rput{43.83}(73.44,87.44){\psellipse[](0,0)(2.21,-2.12)}
\rput(14,60){$\scr{m}$}
\rput(80,64){$\scr{m}$}
\psline[border=0.3,fillstyle=solid](51,110)(57.16,103.84)
\rput(47,120){$\scr{j_z}$}
\psline{|*-}(42,39)(39.17,41.83)
\rput(43,43){$\scr{m}$}
\rput(35,52){$\scr{h_{y}}$}
\psline[fillstyle=solid]{-|}(18.17,62.83)(21,60)
\pspolygon[](17,66)(19,64)(17,62)(15,64)
\rput(21,65){$\scr{R_y}$}
\psline[fillstyle=solid]{|*-}(44.17,36.83)(47,34)
\rput(44,30){$\scr{R^{-1}_y}$}
\rput(38,47){$\scr{R_y}$}
\psline[fillstyle=solid](34.17,46.83)(37,44)
\psline(31,50)(28.18,52.82)
\pspolygon[](27,56)(29,54)(27,52)(25,54)
\rput(31,57){$\scr{R^{-1}_y}$}
\psline[fillstyle=solid]{|-}(23,58)(25.83,55.17)
\rput(49,38){$\scr{m}$}
\pspolygon[](38,45)(40,43)(38,41)(36,43)
\psline[fillstyle=solid](49,32)(51.83,29.17)
\pspolygon[](48,35)(50,33)(48,31)(46,33)
\rput{43.83}(32.44,48.44){\psellipse[](0,0)(2.21,-2.12)}
\psline[fillstyle=solid](11,70)(16.16,64.84)
\psline{|*-}(80,57)(77.17,54.17)
\rput(64,38){$\scr{m}$}
\rput{45}(70.5,47.5){\psellipse[](0,0)(2.21,-2.12)}
\rput(75,46){$\scr{h^{-1}_{x}}$}
\psline[fillstyle=solid]{-|}(56.17,33.17)(59,36)
\pspolygon[](53,32)(55,34)(57,32)(55,30)
\psline[fillstyle=solid](52,29)(54.16,31.16)
\rput(58,28){$\scr{R_x}$}
\psline[fillstyle=solid]{|*-}(82.17,59.17)(85,62)
\rput(88,58){$\scr{R^{-1}_x}$}
\rput(80,52){$\scr{R_x}$}
\psline[fillstyle=solid](72.17,49.17)(75,52)
\psline(69,46)(66.18,43.18)
\pspolygon[](63,42)(65,44)(67,42)(65,40)
\rput(71,40){$\scr{R^{-1}_x}$}
\psline[fillstyle=solid]{|-}(61,38)(63.83,40.83)
\pspolygon[](74,53)(76,55)(78,53)(76,51)
\psline[fillstyle=solid](87,64)(92,69)
\pspolygon[](84,63)(86,65)(88,63)(86,61)
\rput(41,108){$\scr{j_x-m}$}
\rput(58,108){$\scr{j_y+m}$}
\rput(97,67){$\scr{j_y}$}
\rput(5,68){$\scr{j_x}$}
\end{pspicture}
\ea
\end{split}
\ee

The total scalar constraint is then obtained by summing the contributions for $k=1,2$ obtained from the previous expression by index permutations.

This result can now be used to construct explicit solutions or to test the semiclassical limit. 

\section{Conclusions}\label{concl}
We provided a new framework for the cosmological implementation of LQG. This new formulation was aimed to realize a quantum description for an inhomogeneous extension of the Bianchi I model, in which a residual diffeomorphisms invariance held and there was space left to regularize the scalar constraint as in full LQG \cite{qsd}. We outlined how the implementation of a quantization scheme in reduced phase-space was not fit for this purpose. This fact was due to the presence of three independent $U(1)$ gauge symmetries (denoted by $U(1)_i$), each one acting on the integral curves of fiducial vector fields $\omega_i=\partial_i$. The space of invariant states under $U(1)_i$ transformations was made by elements whose $U(1)_i$ quantum numbers were preserved along each curve. The issue of this approach was that such a space is not closed under the action of the scalar constraint, regularized as in \cite{qsd}.   

Henceforth, our new framework has been defined by reversing the order of ``reduction'' and ``quantization'', which means that we projected the kinematical Hilbert space of LQG down to a reduced Hilbert space which captured the degrees of freedom of the extended Bianchi I model. This was done by restricting admissible edges to those ones parallel to fiducial vectors only and by implementing a gauge-fixing procedure for the internal SU(2) symmetry. The former implied that the full diffeomorphisms group was reduced to a proper subgroup, while the latter constituted the most technical part of our analysis. We found the solutions of the gauge-fixing condition by lifting $U(1)_i$ networks to $SU(2)$ ones. This way, we could reconstruct the quantum states describing the extended Bianchi I model out of functions of $SU(2)$ group elements. This feature allowed us to investigate the implications of the original SU(2) invariance, so getting that non-trivial intertwiners  mapping each other different $U(1)_i$ representation. This is the paramount result of our analysis which marked the difference with the reduced quantization scheme. In fact, a true 3-dimensional vertex structure could be realized also in the reduced model. The main consequence was that one could implement the action of the scalar constraint in the reduced model as in full LQG, the only difference being that the triangulation of the spatial manifold had to be replaced by a cubulation. At the same time, the presence of reduced diffeomorphisms allowed to develop certain knot classes over which the scalar constraint could also be consistently regularized. Furthermore, since the volume operator was diagonal, the matrix elements of the scalar constraint can be explicitly computed. For instance, we presented the calculation for a 3-valent vertex structure.         

The analysis of the action of the scalar constraint on the 6-valence vertex and the dynamical implications of the extended Bianchi I model will be the subject of forthcoming investigations. These developments are expected to be highly non-trivial, because the presence of the reduced intertwiners correlates the spin quantum number along different directions already on a kinematical level. In this respect, the construction of a proper semiclassical limit, in which the classical Bianchi I model is inferred, constitutes a tantalizing perspective for testing the proposed quantization procedure. The success of this analysis would qualify  such a scheme as a well-defined quantum picture describing the early Universe in terms of a  discrete geometry, so opening the way to several phenomenological applications. Moreover, it is envisaged for the first time the possibility to test the viability of the techniques developed in LQG (implementation of the scalar constraint \cite{master,Giesel, ar}, development of the semiclassical limit) in a simplified scenario in which the obstructions of the full theory can be overcome.

 This analysis constitutes the first realization of Quantum-reduced Loop Gravity. We applied this framework to the inhomogeneous Bianchi I model, but nothing seems to prevent us for considering other symmetric sectors of the full theory, so increasing the relevance of the proposed procedure and the amount of phenomenological implications which can be extracted.
      
{\acknowledgments
The authors wish to thank T.Thiemann and K.Giesel for useful discussions.
The work of F.C. was supported by funds provided by ``Angelo Della Riccia'' foundation and by the National Science Center under the agreement DEC-2011/02/A/ST2/00294.
The work of E.A. was partially supported by the grant of Polish Narodowe Centrum Nauki nr 2011/02/A/ST2/00300.}

\appendix

\section{SU(2) formulas}\label{app}
The explicit expression of the Wigner function  $d^j_{mn}(\beta)$ \cite{Varsalovi} is:
\be
d^j_{mm'}(\beta)=\sqrt{(j+m)!(j-m)!(j+m')!(j-m')!} \sum_k (-1)^k \frac{(\cos \frac{\beta}{2})^{2j-2k+m-m'}(\sin \frac{\beta}{2})^{2k-m+m'}}{k! (j+m-k)!(j-m'-k)! (m'-m+k)!},
\ee

and the Wigner matrices for an angle $\beta=\frac{\pi}{2}$  are given by
\be
D^{j}_{m,m'}(\alpha,\frac{\pi}{2},\gamma)=(-1)^{m-m'} e^{-i\alpha m-i\gamma m'} \frac{1}{2^j} \sqrt{\frac{(j+m)!(j-m)!}{(j+m')!(j-m')!}}\sum_k(-1)^k \binom{j+m'}{k} \binom{j-m'}{k+m-m'},
\ee
 where the sum over $k$ is such that the argument of the factorials are always bigger than zero. 

\section{Reduced Recoupling }\label{app2}
 
 The standard multiplication of $SU(2)$ holonomies and their recoupling {\it i.e.}

\be
D^{j_1}_{m_1n_1}(g) D^{j_2}_{m_2n_2}(g)=\sum_k C^{km}_{j_1m_1j_2 m_2} D^k_{mn}(g)\;  C^{kn}_{j_1n_1 j_2 n_2} 
\ee  
using the graphical calculus, introduced in \cite{io e antonia}  and based on 3j-symbols related to Clebsch-Gordan coefficients by
\be
C^{j_3m_3}_{j_1m_1 j_2m_2}=(-1)^{j_1-j_2+m_3} \sqrt{d_{j_3}} 
\;
\big(
\begin{array}{ccc}
j_1  & j_2 & j_3 \\
m_1 & m_2 & -m_3
\end{array}
\big),
\ee
can be written as

\be
\begin{array}{c}
\ifx\JPicScale\undefined\def\JPicScale{1}\fi
\psset{unit=\JPicScale mm}
\psset{linewidth=0.3,dotsep=1,hatchwidth=0.3,hatchsep=1.5,shadowsize=1,dimen=middle}
\psset{dotsize=0.7 2.5,dotscale=1 1,fillcolor=black}
\psset{arrowsize=1 2,arrowlength=1,arrowinset=0.25,tbarsize=0.7 5,bracketlength=0.15,rbracketlength=0.15}
\begin{pspicture}(0,0)(25,20)
\psline(12,9)(12,1)
\psline{-<<}(12,5)(5,5)
\psline{<-}(25,5)(18,5)
\psline(12,9)(18,5)
\psline(18,5)(12,1)
\psline(12,20)(12,12)
\psline{-<<}(12,16)(5,16)
\psline{<-}(25,16)(18,16)
\psline(12,20)(18,16)
\psline(18,16)(12,12)
\rput(22,3){$j_1$}
\rput(22,14){$j_2$}
\end{pspicture}
\end{array}
=
\sum_k d_k
\begin{array}{c}
\ifx\JPicScale\undefined\def\JPicScale{1}\fi
\psset{unit=\JPicScale mm}
\psset{linewidth=0.3,dotsep=1,hatchwidth=0.3,hatchsep=1.5,shadowsize=1,dimen=middle}
\psset{dotsize=0.7 2.5,dotscale=1 1,fillcolor=black}
\psset{arrowsize=1 2,arrowlength=1,arrowinset=0.25,tbarsize=0.7 5,bracketlength=0.15,rbracketlength=0.15}
\begin{pspicture}(0,0)(26,10)
\psline(12,9)(12,1)
\psline{-<<}(9,5)(2,2)
\psline{<-}(26,8)(21,5)
\psline(12,9)(18,5)
\psline(18,5)(12,1)
\psline{-<<}(9,5)(2,8)
\psline{<-}(26,2)(21,5)
\rput(3,0){$j_1$}
\rput(2,10){$j_2$}
\psline(9,5)(12,5)
\psline(18,5)(21,5)
\rput(24,9){$j_2$}
\rput(24,1){$j_1$}
\rput(10,3){$k$}
\end{pspicture}
\end{array}
\ee
where the triangle denotes a generic $SU(2)$ group element and the notation with the two kind of arrows is used to distinguish indexes belonging to the vector space $\mathcal{H}^j$ or the dual vector space $\mathcal{H}^{j*}$. The previous expression 
 in the quantum reduced case where we deal with holonomies of the kind \eqref{Dridottamn} becomes 
 
 \be
 \begin {split}
\begin{array}{c}
\ifx\JPicScale\undefined\def\JPicScale{1}\fi
\psset{unit=\JPicScale mm}
\psset{linewidth=0.3,dotsep=1,hatchwidth=0.3,hatchsep=1.5,shadowsize=1,dimen=middle}
\psset{dotsize=0.7 2.5,dotscale=1 1,fillcolor=black}
\psset{arrowsize=1 2,arrowlength=1,arrowinset=0.25,tbarsize=0.7 5,bracketlength=0.15,rbracketlength=0.15}
\begin{pspicture}(0,0)(42,19)
\psline(19,8)(19,0)
\psline{-<<}(19,4)(12,4)
\psline{<-}(32,4)(25,4)
\psline(19,8)(25,4)
\psline(25,4)(19,0)
\psline(19,19)(19,11)
\psline{-<<}(19,15)(12,15)
\psline{<-}(32,15)(25,15)
\psline(19,19)(25,15)
\psline(25,15)(19,11)
\rput(29,2){$j_1$}
\rput(28,13){$j_2$}
\psline{<-|}(42,4)(35,4)
\psline{<-|}(42,15)(35,15)
\psline{|-<<}(9,4)(2,4)
\psline{|-<<}(9,15)(2,15)
\psline(12,16)(12,14)
\psline(32,16)(32,14)
\psline(32,5)(32,3)
\psline(12,5)(12,3)
\rput(6,2){$j_1$}
\rput(6,13){$j_2$}
\rput(38,13){$j_2$}
\rput(38,2){$j_1$}
\end{pspicture}
\end{array}
&=\sum_{k_1k_2k_3} d_{k_1} d_{k_2} d_{k_3}
\begin{array} {c}
\ifx\JPicScale\undefined\def\JPicScale{1}\fi
\psset{unit=\JPicScale mm}
\psset{linewidth=0.3,dotsep=1,hatchwidth=0.3,hatchsep=1.5,shadowsize=1,dimen=middle}
\psset{dotsize=0.7 2.5,dotscale=1 1,fillcolor=black}
\psset{arrowsize=1 2,arrowlength=1,arrowinset=0.25,tbarsize=0.7 5,bracketlength=0.15,rbracketlength=0.15}
\begin{pspicture}(0,0)(75,9)
\psline(38,8)(38,0)
\psline{-<<}(35,4)(28,1)
\psline{<-}(52,7)(47,4)
\psline(38,8)(44,4)
\psline(44,4)(38,0)
\psline{-<<}(35,4)(28,7)
\psline{<-}(52,1)(47,4)
\rput(29,-1){$j_1$}
\rput(28,9){$j_2$}
\psline(35,4)(38,4)
\psline(44,4)(47,4)
\rput(50,8){$j_2$}
\rput(50,0){$j_1$}
\rput(36,2){$k_2$}
\psline{-|}(35,4)(28,1)
\psline{-|}(35,4)(28,7)
\psline{-|}(47,4)(52,7)
\psline{-|}(47,4)(52,1)
\psline{-<<}(64,4)(57,1)
\psline{<-}(75,7)(70,4)
\psline{-<<}(64,4)(57,7)
\psline{<-}(75,1)(70,4)
\rput(58,-1){$j_1$}
\rput(57,9){$j_2$}
\psline(64,4)(67,4)
\psline(67,4)(70,4)
\rput(73,8){$j_2$}
\rput(73,0){$j_1$}
\rput(65,2){$k_3$}
\psline{-|}(64,4)(57,1)
\psline{-|}(64,4)(57,7)
\psline{-<<}(12,4)(5,1)
\psline{<-}(23,7)(18,4)
\psline{-<<}(12,4)(5,7)
\psline{<-}(23,1)(18,4)
\rput(6,-1){$j_1$}
\rput(5,9){$j_2$}
\psline(12,4)(15,4)
\psline(15,4)(18,4)
\rput(21,8){$j_2$}
\rput(21,0){$j_1$}
\rput(13,2){$k_1$}
\psline{-|}(18,4)(23,7)
\psline{-|}(18,4)(23,1)
\end{pspicture}
\end{array}=
\\
&=d_{j_1+j_2}
\begin{array}{c}
\ifx\JPicScale\undefined\def\JPicScale{1}\fi
\psset{unit=\JPicScale mm}
\psset{linewidth=0.3,dotsep=1,hatchwidth=0.3,hatchsep=1.5,shadowsize=1,dimen=middle}
\psset{dotsize=0.7 2.5,dotscale=1 1,fillcolor=black}
\psset{arrowsize=1 2,arrowlength=1,arrowinset=0.25,tbarsize=0.7 5,bracketlength=0.15,rbracketlength=0.15}
\begin{pspicture}(0,0)(58,9)
\psline(28,8)(28,0)
\psline(28,8)(34,4)
\psline(34,4)(28,0)
\rput(22,2){$j_1+j_2$}
\psline{-<<}(7,4)(0,1)
\psline{-<<}(7,4)(0,7)
\rput(1,-1){$j_1$}
\rput(0,9){$j_2$}
\psline{-|}(7,4)(14,4)
\rput(10,2){$j_1+j_2$}
\psline{<-}(58,7)(53,4)
\psline{<-}(58,1)(53,4)
\psline{|-}(46,4)(53,4)
\rput(56,8){$j_2$}
\rput(56,0){$j_1$}
\rput(48,2){$j_1+j_2$}
\psline{-<<}(28,4)(18,4)
\psline(18,5)(18,3)
\psline{<-}(42,4)(34,4)
\psline(42,5)(42,3)
\end{pspicture}
\end{array}
\end{split}
 \ee
where we used  the property of the Clebsch-Gordan $C^{Kk}_{j_1j_1,j_2j_2}$ that is non vanishing only if $K=j_1+j_2$ and $k= j_1+j_2$ graphically given by (remember that in the graph notation we always use 3js):
\be
\ba
\ifx\JPicScale\undefined\def\JPicScale{1}\fi
\psset{unit=\JPicScale mm}
\psset{linewidth=0.3,dotsep=1,hatchwidth=0.3,hatchsep=1.5,shadowsize=1,dimen=middle}
\psset{dotsize=0.7 2.5,dotscale=1 1,fillcolor=black}
\psset{arrowsize=1 2,arrowlength=1,arrowinset=0.25,tbarsize=0.7 5,bracketlength=0.15,rbracketlength=0.15}
\begin{pspicture}(0,0)(34,8)
\psline{<-}(34,7)(29,4)
\psline{<-}(34,1)(29,4)
\psline{>>-}(23,4)(29,4)
\rput(32,8){$j_2$}
\rput(32,0){$j_1$}
\rput(23,2){$j_1+j_2$}
\psline{-|}(29,4)(34,7)
\psline{-|}(29,4)(34,1)
\psline{<-}(11,7)(6,4)
\psline{<-}(11,1)(6,4)
\psline{>>-}(0,4)(6,4)
\rput(9,8){$j_2$}
\rput(9,0){$j_1$}
\rput(1,2){$k$}
\psline{-|}(6,4)(11,7)
\psline{-|}(6,4)(11,1)
\rput(15,4){$=$}
\psline{-|}(29,4)(23,4)
\end{pspicture}
\ea
=\frac{(-1)^{2j_1}}{\sqrt{d_{j_1+j_2}}}
\quad 
\text{and}
 \quad
\ba
\ifx\JPicScale\undefined\def\JPicScale{1}\fi
\psset{unit=\JPicScale mm}
\psset{linewidth=0.3,dotsep=1,hatchwidth=0.3,hatchsep=1.5,shadowsize=1,dimen=middle}
\psset{dotsize=0.7 2.5,dotscale=1 1,fillcolor=black}
\psset{arrowsize=1 2,arrowlength=1,arrowinset=0.25,tbarsize=0.7 5,bracketlength=0.15,rbracketlength=0.15}
\begin{pspicture}(0,0)(38,9)
\psline{-<<}(7,4)(0,1)
\psline{-<<}(7,4)(0,7)
\rput(1,-1){$j_1$}
\rput(0,9){$j_2$}
\psline{->}(7,4)(13,4)
\rput(8,2){$k$}
\psline{-|}(7,4)(0,1)
\psline{-|}(7,4)(0,7)
\rput(16,4){$=$}
\psline{-<<}(32,4)(25,1)
\psline{-<<}(32,4)(25,7)
\rput(26,-1){$j_1$}
\rput(25,9){$j_2$}
\psline{->}(32,4)(38,4)
\rput(33,2){$j_1+j_2$}
\psline{-|}(32,4)(25,1)
\psline{-|}(32,4)(25,7)
\psline{-|}(32,4)(38,4)
\end{pspicture}
\ea
=\frac{(-1)^{2j_2}}{\sqrt{d_{j_1+j_2}}}
\ee

The same result can be obtained remembering that
\be
\ba
\ifx\JPicScale\undefined\def\JPicScale{1}\fi
\psset{unit=\JPicScale mm}
\psset{linewidth=0.3,dotsep=1,hatchwidth=0.3,hatchsep=1.5,shadowsize=1,dimen=middle}
\psset{dotsize=0.7 2.5,dotscale=1 1,fillcolor=black}
\psset{arrowsize=1 2,arrowlength=1,arrowinset=0.25,tbarsize=0.7 5,bracketlength=0.15,rbracketlength=0.15}
\begin{pspicture}(0,0)(35,10)
\psline{-<<}(9,5)(2,2)
\psline{-<<}(9,5)(2,8)
\rput(3,0){$j_1$}
\rput(2,10){$j_2$}
\psline{->}(9,5)(13,5)
\rput(10,3){$k$}
\psline{<-}(40,8)(35,5)
\psline{<-}(40,2)(35,5)
\psline{>>-}(30,5)(35,5)
\rput(38,9){$j_2$}
\rput(38,1){$j_1$}
\rput(33,3){$k$}
\rput(22,5){$=(-1)^{2k}$}
\end{pspicture}
\ea
\ee
and the coherent state property $|j,j>=|\frac{1}{2},\frac{1}{2}>^{\bigotimes 2j}$ that graphically implies:

\be
\ba
\ifx\JPicScale\undefined\def\JPicScale{1}\fi
\psset{unit=\JPicScale mm}
\psset{linewidth=0.3,dotsep=1,hatchwidth=0.3,hatchsep=1.5,shadowsize=1,dimen=middle}
\psset{dotsize=0.7 2.5,dotscale=1 1,fillcolor=black}
\psset{arrowsize=1 2,arrowlength=1,arrowinset=0.25,tbarsize=0.7 5,bracketlength=0.15,rbracketlength=0.15}
\begin{pspicture}(-9,0)(35,11)
\psline{<-|}(-1,4)(-8,4)
\psline{<-|}(-1,10)(-8,10)
\rput(0,8){$j_2$}
\rput(0,2){$j_1$}
\psline{<-}(34,10)(29,7)
\psline{<-}(34,4)(29,7)
\psline{|-}(22,7)(29,7)
\rput(32,11){$j_2$}
\rput(32,3){$j_1$}
\rput(24,5){$j_1+j_2$}
\rput(10,7){$=\sqrt{d_{j_1+j_2}}$}
\end{pspicture}
\ea
\quad \quad
\text{and}
 \quad
\begin {array} {c}
\ifx\JPicScale\undefined\def\JPicScale{1}\fi
\psset{unit=\JPicScale mm}
\psset{linewidth=0.3,dotsep=1,hatchwidth=0.3,hatchsep=1.5,shadowsize=1,dimen=middle}
\psset{dotsize=0.7 2.5,dotscale=1 1,fillcolor=black}
\psset{arrowsize=1 2,arrowlength=1,arrowinset=0.25,tbarsize=0.7 5,bracketlength=0.15,rbracketlength=0.15}
\begin{pspicture}(-9,0)(38,12)
\psline{|-<<}(-1,4)(-8,4)
\psline{|-<<}(-1,10)(-8,10)
\rput(1,2){$j_1$}
\rput(1,8){$j_2$}
\psline{-<<}(29,7)(22,4)
\psline{-<<}(29,7)(22,10)
\rput(23,2){$j_1$}
\rput(22,12){$j_2$}
\psline{-|}(29,7)(36,7)
\rput(32,5){$j_1+j_2$}
\rput(11,7){$=\sqrt{d_{j_1+j_2}}$}
\rput(38,12){}
\end{pspicture}
\end {array}
\ee

In the case in which the two groups elements go in opposite directions we have instead

\be
\ba
\ifx\JPicScale\undefined\def\JPicScale{1}\fi
\psset{unit=\JPicScale mm}
\psset{linewidth=0.3,dotsep=1,hatchwidth=0.3,hatchsep=1.5,shadowsize=1,dimen=middle}
\psset{dotsize=0.7 2.5,dotscale=1 1,fillcolor=black}
\psset{arrowsize=1 2,arrowlength=1,arrowinset=0.25,tbarsize=0.7 5,bracketlength=0.15,rbracketlength=0.15}
\begin{pspicture}(0,0)(42,19)
\psline(19,19)(19,11)
\psline{-<<}(19,15)(12,15)
\psline{<-}(32,15)(25,15)
\psline(19,19)(25,15)
\psline(25,15)(19,11)
\rput(28,13){$j_2$}
\psline{<-|}(2,5)(9,5)
\psline{<-|}(42,15)(35,15)
\psline{|-<<}(35,5)(42,5)
\psline{|-<<}(9,15)(2,15)
\psline(12,16)(12,14)
\psline(32,16)(32,14)
\rput(38,3){$j_1$}
\rput(6,13){$j_2$}
\rput(38,13){$j_2$}
\rput(6,3){$j_1$}
\psline(19,9)(19,1)
\psline{->}(19,5)(12,5)
\psline{>>-}(32,5)(25,5)
\psline(19,9)(25,5)
\psline(25,5)(19,1)
\rput(28,3){$j_1$}
\psline(12,6)(12,4)
\psline(32,6)(32,4)
\end{pspicture}
\ea
=
\frac{(d_{j_2-j_1})^3}{(d_{j_2})^2}
\ba
\ifx\JPicScale\undefined\def\JPicScale{1}\fi
\psset{unit=\JPicScale mm}
\psset{linewidth=0.3,dotsep=1,hatchwidth=0.3,hatchsep=1.5,shadowsize=1,dimen=middle}
\psset{dotsize=0.7 2.5,dotscale=1 1,fillcolor=black}
\psset{arrowsize=1 2,arrowlength=1,arrowinset=0.25,tbarsize=0.7 5,bracketlength=0.15,rbracketlength=0.15}
\begin{pspicture}(0,0)(58,9)
\psline(28,8)(28,0)
\psline(28,8)(34,4)
\psline(34,4)(28,0)
\rput(23,2){$\scr{j_2-j_1}$}
\psline{->}(7,4)(0,1)
\psline{-<<}(7,4)(0,7)
\rput(1,-1){$j_1$}
\rput(0,9){$j_2$}
\psline{-|}(7,4)(14,4)
\rput(10,2){$\scr{j_2-j_1}$}
\psline{<-}(58,7)(53,4)
\psline{>>-}(58,1)(53,4)
\psline{|-}(46,4)(53,4)
\rput(56,8){$j_2$}
\rput(56,0){$j_1$}
\rput(48,2){$\scr{j_2-j_1}$}
\psline{-<<}(28,4)(18,4)
\psline(18,5)(18,3)
\psline{<-}(42,4)(34,4)
\psline(42,5)(42,3)
\end{pspicture}
\ea
\ee
because we can recouple the lines obtaing a sum over allowed spins, but in the reduced space ${}^\mathcal{G}\mathcal{H}^R$ the only non vanishing terms are produced by the Clebsch-Gordan (3j) of the kind:
\be
\ba
\ifx\JPicScale\undefined\def\JPicScale{1}\fi
\psset{unit=\JPicScale mm}
\psset{linewidth=0.3,dotsep=1,hatchwidth=0.3,hatchsep=1.5,shadowsize=1,dimen=middle}
\psset{dotsize=0.7 2.5,dotscale=1 1,fillcolor=black}
\psset{arrowsize=1 2,arrowlength=1,arrowinset=0.25,tbarsize=0.7 5,bracketlength=0.15,rbracketlength=0.15}
\begin{pspicture}(0,0)(34,8)
\psline{<-}(34,7)(29,4)
\psline{<-}(34,1)(29,4)
\psline{>>-}(23,4)(29,4)
\rput(32,9){$j_2-j_1$}
\rput(32,0){$j_1$}
\rput(23,2){$j_2$}
\psline{-|}(29,4)(34,7)
\psline{-|}(29,4)(34,1)
\psline{-|}(29,4)(23,4)
\end{pspicture}
\ea
=
\frac{(-1)^{2 j_1}}{\sqrt{d_{j_2}}}
\quad
\ba
\ifx\JPicScale\undefined\def\JPicScale{1}\fi
\psset{unit=\JPicScale mm}
\psset{linewidth=0.3,dotsep=1,hatchwidth=0.3,hatchsep=1.5,shadowsize=1,dimen=middle}
\psset{dotsize=0.7 2.5,dotscale=1 1,fillcolor=black}
\psset{arrowsize=1 2,arrowlength=1,arrowinset=0.25,tbarsize=0.7 5,bracketlength=0.15,rbracketlength=0.15}
\begin{pspicture}(15,0)(38,9)
\psline{-<<}(32,4)(25,1)
\psline{-<<}(32,4)(25,7)
\rput(26,-1){$j_1$}
\rput(25,9){$j_2-j_1$}
\psline{->}(32,4)(38,4)
\rput(33,2){$j_2$}
\psline{-|}(32,4)(25,1)
\psline{-|}(32,4)(25,7)
\psline{-|}(32,4)(38,4)
\end{pspicture}
\ea
=\frac{(-1)^{2j_1+2j_2}}{\sqrt{d_{j_2}}}
\ee

\end{document}